\documentclass[letterpaper,apj]{emulateapj}
\pdfoutput=1
\usepackage{hyperref}
\usepackage{amsmath,amstext}
\usepackage[T1]{fontenc}
\usepackage[figure,figure*]{hypcap}

\shorttitle{The Spectral Evolution of Cool WDs}
\shortauthors{Blouin et al.}

\begin{document}

\submitted{Accepted for publication in The Astrophysical Journal}

\title{A New Generation of Cool White Dwarf Atmosphere Models.
  IV. Revisiting the Spectral Evolution of Cool White Dwarfs}

\author{S. Blouin\altaffilmark{1}}
\author{P. Dufour\altaffilmark{1}}
\author{C. Thibeault\altaffilmark{1}}
\author{N.~F. Allard\altaffilmark{2,3}}

\altaffiltext{1}{D\'epartement de Physique, Universit\'e de Montr\'eal, Montr\'eal, QC H3C 3J7, Canada; sblouin@astro.umontreal.ca, dufourpa@astro.umontreal.ca}
\altaffiltext{2}{GEPI, Observatoire de Paris, Universit\'e PSL, CNRS, UMR 8111, 61 avenue de l'Observatoire, F-75014 Paris, France}
\altaffiltext{3}{Sorbonne Universit\'e, CNRS, UMR 7095, Institut d'Astrophysique de Paris, 98bis boulevard Arago, F-75014 Paris, France}

\begin{abstract}
  As a result of competing physical mechanisms, the atmospheric composition of white
  dwarfs changes throughout their evolution, a process known as spectral evolution. 
  Because of the ambiguity of their atmospheric
  compositions and the difficulties inherent to the modeling of their dense atmospheres,
  no consensus exists regarding the spectral evolution of cool white dwarfs 
  ($T_{\rm eff}<6000\,{\rm K}$).
  In the previous papers of this series, we presented and observationally validated
  a new generation of cool white dwarf atmosphere models that include all 
  the necessary constitutive physics to accurately model those objects.
  Using these new models and a homogeneous sample of 501 cool white dwarfs, we revisit
  the spectral evolution of cool white dwarfs. Our sample includes all 
  spectroscopically identified white dwarfs cooler than 8300\,K for which a parallax
  is available in {\it Gaia} DR2 and photometric observations are available in
  Pan-STARRS1 and 2MASS.
  Except for a few cool carbon-polluted objects, our models allow an excellent fit
  to the spectroscopic and photometric observations of all objects included in our
  sample. 
  We identify a decrease of the ratio of hydrogen to helium-rich objects between
  7500\,K and 6250\,K, which we interpret as the signature of convective mixing.
  After this decrease, hydrogen-rich objects become more abundant up to 5000\,K.
  This puzzling increase, reminiscent of the non-DA gap, has yet to be explained.
  At lower temperatures, below 5000\,K, hydrogen-rich white dwarfs become rarer,
  which rules out the
  scenario according to which accretion of hydrogen from the interstellar medium
  dominates the spectral evolution of cool white dwarfs.
\end{abstract}

\section{Introduction}
White dwarf stars are compact stellar remnants condemned to a slow cooling that
extends over billions of years. One of the fundamental properties of white dwarfs is their
intense surface gravity ($\log g=8$ on average). This strong gravitational field implies 
that heavy elements are expected to sink towards the core of the white dwarf and that
light elements should float high up in the atmosphere \citep{schatzman1945theorie}.
If gravitational settling was the only process influencing the chemical composition
of white dwarf atmospheres, then all white dwarfs would show pure hydrogen atmospheres
(i.e., only the DA and DC spectral types would exist).

However, we know that at least $\sim 20\%$
of white dwarfs have helium-rich atmospheres \citep{giammichele2012know,kepler2015new},
that many have metal-polluted atmospheres (DQs and DZs) and that some even have
carbon-dominated \citep[Hot DQs,][]{dufour2007white}
or oxygen-dominated atmospheres \citep[DS\footnote{DS is the proposed classification for oxygen-rich objects \citep[see also Table~8 of][]{leggett2018distant}.},][]{kepler2016white}. These simple
observational facts imply that additional mechanisms can alter the atmospheric
composition of a white dwarf. For instance, a white dwarf atmosphere can undergo
simple diffusion \citep{muchmore1984diffusion,paquette1986diffusion,koester2009accretion},
convective dilution \citep{fontaine1987recent,macdonald1991how,rolland2018spectral},
convective mixing \citep{koester1976convective,vauclair1977elements,dantona1979white,rolland2018spectral},
convective dredge-up from the core \citep{pelletier1986carbon},
radiative levitation \citep{chayer1995radiative},
accretion from the interstellar medium \citep{dupuis1993study,koester2006accretion} and
accretion of rocky planetesimals \citep{graham1990infrared,jura2003tidally,farihi2010rocky}.
As those mechanisms change in importance during the evolution of white dwarfs, their
surface compositions also change as a function of decreasing effective temperature.
This phenomenon, known as spectral evolution, is one of the most studied aspects of
white dwarf evolution.

The most spectacular feature of the spectral evolution of white dwarfs is probably
the existence of a so-called DB gap between $T_{\rm eff} \approx 45\,000\,{\rm K}$ and
$30\,000\,{\rm K}$ \citep{liebert1986origin,liebert1986temperatures}, where helium-rich
white dwarfs are significantly less abundant than at higher or lower temperatures \citep{eisenstein2006hot}.
This deficiency has been explained as the result of a hydrogen float-up
process at the blue edge of the gap and a convective dilution process below the red edge
\citep{fontaine1987recent,macdonald1991how}.

Going down the temperature scale, the next striking feature in the spectral evolution
of white dwarfs is the increase of the ratio of non-DA to DA objects below
$T_{\rm eff} \approx 12\,000\,{\rm K}$ \citep{sion1984implications,tremblay2008ratio}. 
This observation is the consequence of
the convective mixing of a thin hydrogen layer with the thick helium envelope underneath
that can transform a hydrogen-rich atmosphere into a helium-rich atmosphere
\citep{koester1976convective,vauclair1977elements,dantona1979white,rolland2018spectral}. 
The lower the temperature of the star, the deeper the convection zone extends into the 
white dwarf. Therefore, objects with a thicker hydrogen layer turn into helium-rich stars 
later in their evolution \citep{tassoul1990evolutionary,bergeron1997chemical}.

At lower temperatures, however, our understanding of the spectral evolution of white dwarfs
is more limited. In fact, results obtained in the last two decades point in conflicting
directions. On the one hand, analyses performed by \cite{bergeron1997chemical,
  bergeron2001photometric} and \cite{kilic2006cool,kilic2010detailed} have revealed
that the ratio of hydrogen-rich to helium-rich stars is greatly enhanced between
5000 and 6000\,K. Due to its deficiency of helium-rich objects, this temperature
range was termed the non-DA gap. The existence of this gap is usually understood as
the consequence of processes that transform a significant fraction of
helium-rich stars into hydrogen-rich stars (near 6000\,K) and that later transform
hydrogen-rich objects into helium-rich objects (near 5000\,K). So far, no
consistent physical explanation for the existence of the non-DA gap has been proposed
\citep{hansen1999cooling,malo1999physical,bergeron2001photometric}.

On the other hand, results obtained by \cite{kowalski2006found} and
\cite{kilic2009spitzer,kilic2009near} suggest that the overabundance of hydrogen-rich object
is not limited to the 5000$-$6000\,K range and that it extends all the way to the coolest
objects at $T_{\rm eff} \approx 4000\,{\rm K}$. In fact, they find that almost every single
object in their samples is hydrogen-rich for $T_{\rm eff}<5000\,{\rm K}$. The overabundance
of hydrogen-rich stars between 5000 and 6000\,K would therefore not be a gap but rather a portion
of a continuous process that irreversibly transforms helium-rich atmosphere into hydrogen-rich
atmospheres. \cite{kowalski2006phd} has tentatively suggested that this transformation
could be due to the accretion of hydrogen from the interstellar medium.

The fact that the atmospheric composition of the coolest white dwarfs is still open to
debate is concerning for white dwarf age dating. As white dwarfs cool down monotonically
and as their cooling rate can be precisely modeled
\citep{hansen1999cooling,fontaine2001potential,renedo2010new},
they can be very accurate cosmic clocks
\citep[for applications to various stellar populations, see][]{oswalt1996lower,garcia2010white,kalirai2012age,tremblay2014white}.
However, in order to use a white dwarf as a cosmochronometer,
a precise determination of its atmospheric parameters
is required. In particular, it is important to know its
atmospheric composition, since helium-rich and hydrogen-rich atmospheres do not have the
same opacities and therefore have different cooling rates. For instance, for a white dwarf with
$T_{\rm eff}=4000\,{\rm K}$ and $\log g=8$, a mistake on the atmospheric composition can
lead to an error of $\sim 1\,$Gyr on its cooling age
\citep[][assuming that the white dwarf evolves with a constant atmospheric composition]{fontaine2001potential}.
In this context, knowing the atmospheric composition of the coolest (and thus oldest)
white dwarfs becomes an even more pressing problem.

There are two main reasons that explain the discrepancies between studies that conclude to
the existence of the non-DA gap and those that suggest a continuous increase of the
hydrogen-rich to helium-rich ratio. The first one is that the samples of
\cite{kowalski2006found} and \cite{kilic2009spitzer,kilic2009near} are limited to
DA and DC white dwarfs. Therefore, they completely ignore the existence of DQs and DZs, which
have helium-rich atmospheres \citep[for limits on the hydrogen abundance in the atmospheres
  of DQs and DZs, see][]{dufour2005detailed,dufour2007spectral}. For this reason, their
samples are strongly biased towards hydrogen-rich stars. In particular, we can be confident
that helium-rich white dwarfs do exist below 5000\,K, since DQs and DZs are found
at those temperatures.

The second reason that explains the discrepancy between both sets of studies is related to
the fact that both helium-rich and hydrogen-rich stars become DCs below $\approx 5000\,{\rm K}$.
In fact, the thermal energy becomes too small to excite the atomic states that are required to produce
hydrogen or helium spectral lines in the visible and most white dwarfs with $T_{\rm eff} \lesssim 5000\,{\rm K}$
thus show a flat, featureless spectrum. This implies that the chemical composition of the atmospheres
of such stars must be derived solely from the photometric observations (since no useful
information can be retrieved from the spectroscopy).
Atmosphere models are used to fit the spectral energy distribution (SED)
assuming both hydrogen-rich and helium-rich models and the best solution indicates
the most likely composition of the atmosphere. As this process depends on
a detailed fit of the SED, small differences between different
sets of atmosphere models can lead to different solutions. The same star can be classified
as having a hydrogen-rich atmosphere using one set of models and as having a helium-rich
atmosphere using another set of models. This is precisely what happens here.
Compared to the models of
\citet[used in \citealt{bergeron1997chemical,bergeron2001photometric,kilic2006cool,kilic2010detailed}]{bergeron1995new}, 
the models employed in \cite{kowalski2006found} and \cite{kilic2009spitzer,kilic2009near}
include numerous improvements for the treatment of nonideal effects in the dense atmospheres of
cool white dwarfs. These improved models predict that cool helium-rich white dwarfs have SEDs
that are very close to that of blackbodies. As this is not observed, cool white dwarfs are almost
all classified as hydrogen-rich \citep{kowalski2006phd}.

This paper aims at fixing the issues that tarnished previous analyses of the spectral
evolution of cool white dwarfs. On the theoretical front, our analysis is based on
state-of-the-art atmosphere models that take into account all high-density effects relevant
for the modeling of the atmospheres of cool white dwarfs \citep[][hereafter Paper I]{blouin2018model}.
These new models have the advantage of being also applicable to metal-polluted white dwarfs.
Not only is this useful for modeling all cool white dwarfs (and not only pure hydrogen and pure
helium objects), but it is also an excellent way of observationally validating the input physics
of our models. Indeed, as metal-polluted white dwarfs (DQs and DZs) are the only ones to still show spectral
features in the visible below $T_{\rm eff} \approx 5000\,{\rm K}$, they represent a unique opportunity of
testing our models against observations. In contrast, as there is no challenge in fitting the
flat spectrum of DC white dwarfs, they cannot be used to discriminate between good and poor models. Using
this insight, we tested our models against some of the most challenging cool metal-polluted
white dwarfs (see Paper I and \citealt{blouin2018dzcia,blouin2019wd2356}, hereafter Papers II and III).
In every case, we showed that our models can accurately reproduce the complex spectral features
observed in those objects, which validates their constitutive physics.

On the observational front, the analysis presented in this paper is based on the largest (501 objects)
homogeneous sample of cool white dwarfs ever studied. The homogeneity of our sample is guaranteed
by the fact that all our fits are based on data obtained
from the same three surveys (Pan-STARRS1, 2MASS and \textit{Gaia} DR2).
Additionally, no discrimination on the spectral types was made during the selection of the sample,
which ensures a good representation of the diversity of the chemical compositions of cool white
dwarf atmospheres.

Section~\ref{sec:methodology} details our methodology, including the selection of the sample,
the model atmosphere code and the fitting procedures. In Section~\ref{sec:results}, we present
the solutions of our fits and analyze in detail some interesting objects. In particular, we
turn our attention to the so-called peculiar non-DAs---stars for which the photometric
and spectroscopic observations suggest conflicting atmospheric compositions \citep{bergeron1997chemical}---and
to DQpec white dwarfs---carbon-polluted objects with distorted Swan bands \citep{hall2008c2}. 
After a careful analysis of the biases that affect our sample,
Section~\ref{sec:evolution} presents our conclusions on the spectral evolution of cool white dwarfs.
Finally, our main findings are summarized in Section~\ref{sec:conclusion}.

\section{Methodology}
\label{sec:methodology}
\subsection{Sample selection}
\label{sec:sample_selection}
We defined the selection criteria of our sample so that a maximum
number of white dwarfs could be included, while ensuring that the
observations used for the photometric fits are homogeneous.
Six criteria must be satisfied for an object to be part of our sample.
\begin{enumerate}
\item It must have a parallax measurement from the \textit{Gaia} DR2
  \citep{prusti2016gaia,brown2018gaia}. The parallax is important not
  only for measuring the mass of the white dwarf, but also for
  identifying unresolved binary systems.
\item Each object must have $grizy$ photometry from
  the Panoramic Survey Telescope and Rapid Response System
  \citep[Pan-STARRS1,][]{chambers2016panstarrs}.
\item For each object, there must be at least $J$ photometry
  (ideally $H$ and $K$ also) in the Two Micron All Sky Survey (2MASS).
\item A spectrum must be available for each object, so that the
  appropriate atmospheric composition can be assumed in our models. In particular,
  the presence of H$\alpha$ is useful for identifying hydrogen-rich
  atmospheres and metal lines and C$_2$ Swan bands are used to fix the
  amount of heavy elements in the model atmospheres of DZ and DQ white dwarfs,
  respectively.
\item Stars that are part of a known unresolved binary system were rejected.
\item Only objects cooler than 8300\,K were retained. This upper limit is motivated
  by our wish to verify the existence of the non-DA gap. To do so, we need to compare
  the abundance of hydrogen-rich stars above and below the blue edge of this gap
  at $T_{\rm eff} \approx 6000\,$K. The 8300\,K value was chosen as it allows
  to compute the fraction of hydrogen-rich stars as a function of effective temperature
  for 500\,K bins up to a bin centered at 8000\,K.
\end{enumerate}

We relied on the Montreal White Dwarf Database
\citep[MWDD,][]{dufour2016montreal} to identify objects that match these
criteria. In total, we found 503 stars that satisfy all six criteria.
The bulk of these objects are part of the 
\cite{limoges2015physical} sample (292 objects) and the rest comes
from a number of other studies
\citep{bergeron1997chemical,bergeron2001photometric,bergeron2005interpretation,
  putney1997surveying,
  kawka2006spectroscopic,kawka2012study,
  subasavage2007solar,subasavage2008solar,subasavage2009solar,subasavage2017solar,
  kilic2010detailed,
  giammichele2012know,
  sayres2012multi,
  kleinman2013sdss,
  gianninas2015ultracool,
  kepler2015new,kepler2016new}.
Among the 503 objects initially selected, two of them (GJ~1221 and GJ~1228)
were rejected because no appropriate models are available at the moment to 
properly model their atmospheres. These two white dwarfs,
classified as DXP stars, are characterized by
broad, unidentified absorption features and a very large magnetic field
\citep[$B>100\,{\rm MG}$,][]{berdyugin1999polarization,putney1995off}.
Therefore, our final sample contains 501 objects, which are listed in
Tables~\ref{tab:astrometry} (astrometric data) and \ref{tab:photometry} (photometric data).

Due to the relatively small limiting magnitude of 2MASS, requiring each
object to be in 2MASS considerably reduces the size of our sample.
Nevertheless, this constraint is important since observations in
the infrared are necessary to detect collision-induced absorption (CIA)
from molecular hydrogen, which is very useful to constrain the hydrogen
abundance. This is a particularly important parameter to determine if we
want to get an accurate picture of the spectral evolution of cool DC white
dwarfs.

\begin{deluxetable*}{llllll}
  \tabletypesize{\footnotesize}
  \tablecaption{Astrometry of objects included in our sample \label{tab:astrometry}}
  \tablehead{PSO\tablenotemark{a} & MWDD object ID & Gaia object ID &
    R.A. (J2015.5) & Decl. (J2015.5) &  $\pi$ (mas)}
  \startdata
 J000224.474+635745.512   &  2MASS J00022257+6357443 &   431635455820288128 &     0.6033724 &    \phm{$-$}63.9627281 &    38.080(0.079) \\
 J000410.478$-$034008.751 &                 PHL 2595 &  2447889401738675072 &     1.0439384 &           $-$3.6691769 &    21.076(0.100) \\
 J000720.831+123018.402   &              WD 0004+122 &  2766234439302571904 &     1.8371872 &    \phm{$-$}12.5048777 &    57.308(0.114) \\
 J000728.917+340342.227   &                 NLTT 301 &  2875903332533220992 &     1.8705266 &    \phm{$-$}34.0618834 &    29.336(0.099) \\
 J000754.487+394731.294   &              LP 240$-$30 &   383108338321272448 &     1.9774884 &    \phm{$-$}39.7919426 &    29.038(0.064) \\
 J000935.139+310840.332   &                   EGGR 1 &  2861792754354276352 &     2.3959624 &    \phm{$-$}31.1440912 &    18.638(0.077) \\
 J001122.399+424038.082   &                     GD 5 &   384636109728592768 &     2.8432990 &    \phm{$-$}42.6770566 &    42.714(0.045) \\
 J001214.188+502514.381   &                  GJ 1004 &   395234439752169344 &     3.0583199 &    \phm{$-$}50.4200515 &    91.983(0.029) \\
 J001412.352$-$131109.057 &                  GJ 3016 &  2418116963320446720 &     3.5508822 &          $-$13.1867135 &    53.812(0.070) \\
 J001737.914$-$051650.624 &              LP 644$-$81 &  2443826805441462656 &     4.4082578 &          $-$5.28121904 &    21.405(0.153) 
 \enddata
 \tablenotetext{a}{Pan-STARRS object name}
 \tablecomments{Table~\ref{tab:astrometry} is published in its entirety in the machine-readable format. A portion is shown here for guidance regarding its form and content.}
\end{deluxetable*}

\begin{deluxetable*}{lccccccccc}
  \tabletypesize{\footnotesize}
  \tablecaption{Spectral type and photometric data \label{tab:photometry}}
  \tablehead{PSO & Spectral type
    & $g$ & $r$ & $i$ & $z$ & $y$ & $J$ & $H$ & $K$ }
  \startdata
 J000224.474+635745.512   &       DC &  17.63 &  16.99 &  16.73 &  16.63 &  16.57 &   15.80 &   15.57 &   15.51 \\
 J000410.478$-$034008.751 &       DA &  16.91 &  16.75 &  16.73 &  16.76 &  16.79 &   16.12 &   15.89 &   15.44 \\
 J000720.831+123018.402   &       DC &  16.77 &  16.25 &  16.02 &  15.93 &  15.92 &   15.08 &   15.10 &   14.90 \\
 J000728.917+340342.227   &       DC &  17.61 &  17.23 &  17.12 &  17.05 &  17.08 &   16.39 &   16.25 &   15.83 \\
 J000754.487+394731.294   &       DC &  17.25 &  16.53 &  16.23 &  16.12 &  16.06 &   15.18 &   14.85 &   14.65 \\
 J000935.139+310840.332   &       DC &  16.81 &  16.74 &  16.85 &  16.94 &  17.01 &   16.45 &   16.19 &     --  \\
 J001122.399+424038.082   &       DA &  15.42 &  15.27 &  15.23 &  15.28 &  15.31 &   14.54 &   14.35 &   14.39 \\
 J001214.188+502514.381   &      DAH &  14.52 &  14.27 &  14.20 &  14.21 &  14.22 &   13.49 &   13.25 &   13.19 \\
 J001412.352$-$131109.057 &      DAH &  16.12 &  15.76 &  15.62 &  15.59 &  15.60 &   14.81 &   14.55 &   14.63 \\
 J001737.914$-$051650.624 &       DA &  17.92 &  17.52 &  17.35 &  17.32 &  17.27 &   16.48 &   16.35 &     --  
 \enddata
  \tablecomments{Table~\ref{tab:photometry} is published in its entirety in the machine-readable format.}
\end{deluxetable*}

\subsection{Atmosphere models}
Our model atmosphere code is described at length in Paper I.
It is based on the code described in
\cite{bergeron1995new} and \cite{dufour2005detailed,dufour2007spectral}, but also
includes a number of nonideal high-density effects that arise in the dense
atmospheres of cool white dwarfs. In particular, continuum opacities are
corrected for collective interactions
\citep{iglesias2002density,rohrmann2018rayleigh}, an ab initio equation
of state for hydrogen and helium is assumed \citep{becker2014ab},
the pressure ionization of helium is modeled using the ab initio calculations
of \cite{kowalski2007equation},
an accurate description of the pressure broadening of Ly$\alpha$ is
included \citep{kowalski2006found},
the H$_2-$He CIA profiles are corrected for many-body collisions
\citep{blouin2017pressure} and CIA from He$-$He$-$He interactions is included
\citep{kowalski2014infrared}. Moreover, our code includes significant
improvements relevant for the accurate modeling of cool metal-polluted atmospheres,
such as accurate line profiles for important heavy element lines
\citep[Paper III;][]{allard2014caii,allard2014temperature,
  allard2016theoretical,allard2016asymmetry,allard2018line,blouin2019line},
a state-of-the-art treatment of the nonideal ionization equilibrium
of C, Ca, Fe, Mg and Na (Paper I), and the distortion of
the C$_2$ Swan bands in DQpec stars \citep{kowalski2010origin}.
Note also that we now include the C$_2$ Swan bands opacity using a line-by-line approach
with the linelist of \cite{hornkohl2005diatomic}. This approach noticeably improves
our fit to the shape
of the Swan bands compared to the just overlapping line approximation \citep{zeidler1982just}
implemented in the models of \cite{dufour2005detailed}.

\subsection{Fitting procedures}
\label{sec:fitting_procedure}
\subsubsection{DAs}
\label{sec:DA}
The fundamental parameters of DA white dwarfs are obtained with the
photometric technique \citep{bergeron1997chemical}.
The solid angle $\pi (R/D)^2$ and
$T_{\rm eff}$ are found by fitting the model fluxes to the observed
SED with the Levenberg--Marquardt algorithm.
Since $D$ is known from the \textit{Gaia} parallax measurement, the white
dwarf radius $R$ can be computed from the solid angle. The mass and the surface
gravity of the white dwarf are then found with the evolutionary models
of \cite{fontaine2001potential}.
Note that for all DA stars in our sample we assume a pure hydrogen atmosphere.
  This assumption is very common in the literature
  \citep[e.g.][]{bergeron2001photometric,gianninas2011spectroscopic,limoges2015physical}
  and is supported by the comparison between synthetic and observed Balmer lines profiles.
  For cool white dwarfs, Balmer lines become shallower and wider when the
  H/He ratio is decreased. 
  While it is not possible to distinguish between Balmer lines
  created in a pure hydrogen atmosphere and an atmosphere with $\log\,{\rm H/He}=2$, the differences
  between a pure hydrogen atmosphere and a helium-dominated atmosphere ($\log\,{\rm H/He}<0$) are
  obvious.
  As our sample is largely based on the sample of \cite{limoges2015physical} and as
  their spectroscopic analysis shows no evidence of helium-dominated DAs, we conclude that
  we can safely assume that all DAs in our sample are hydrogen-rich.
  There remains the question of whether they simply have a hydrogen-dominated atmosphere
  or a pure hydrogen atmosphere, but since we cannot distinguish between both possibilities
  we assume a pure hydrogen composition for simplicity. Note that this assumption
  does not interfere with the main purpose of this work, since we only need to know
  which stars are hydrogen-rich and which ones are helium-rich.

\subsubsection{DCs}
\label{sec:procedure_DC}
\begin{figure*}
    \includegraphics[width=\linewidth]{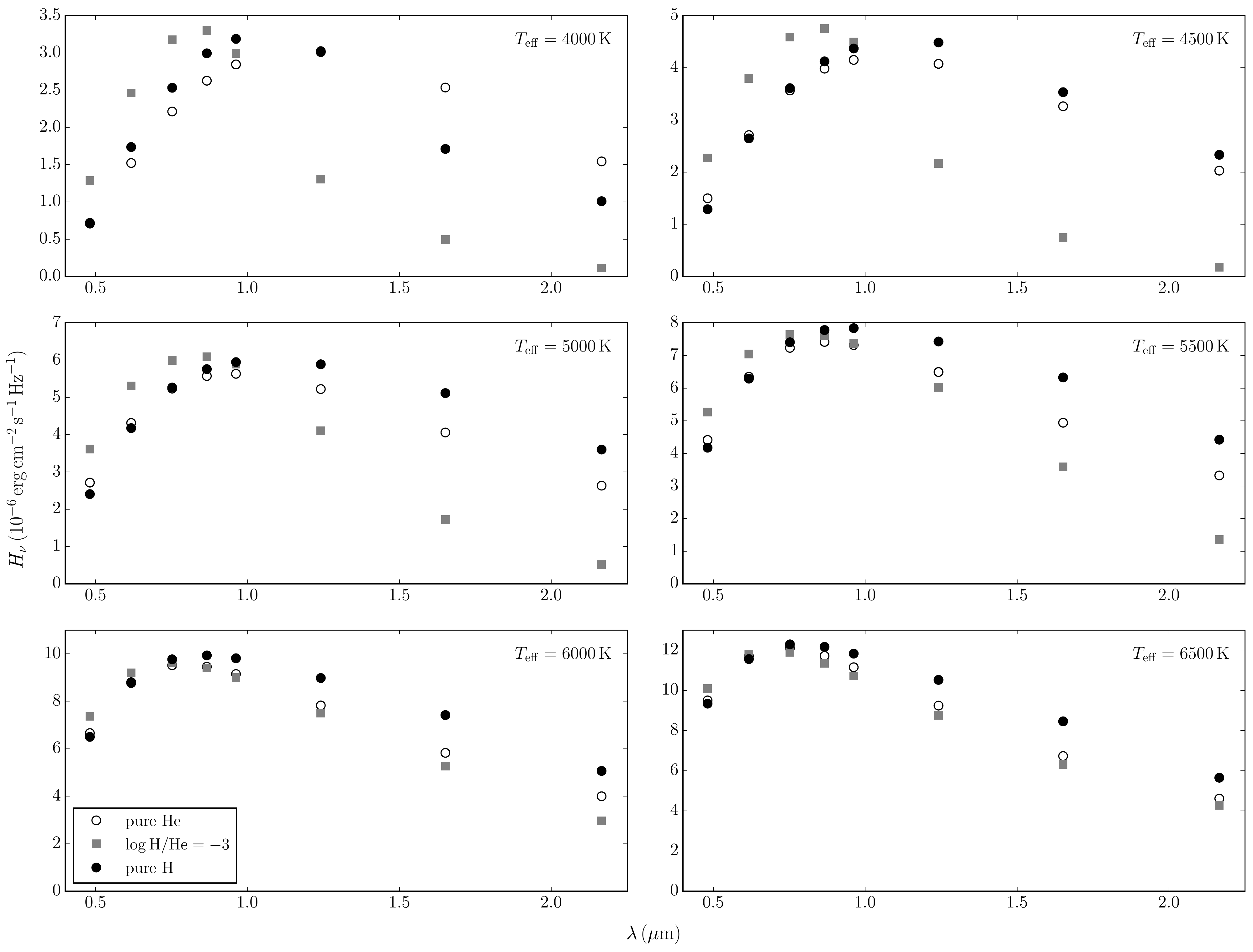}
    \caption{Synthetic Pan-STARRS and 2MASS photometry obtained with our
      model atmosphere code. Each panel corresponds to a different effective
      temperature and each symbol represents a different atmospheric
      composition (see legend). In all cases, a surface gravity of $\log g=8$
      is assumed.}
  \label{fig:model_grid}
\end{figure*}

The fitting procedure for DC stars is identical to the procedure described
above for DAs, except that the atmospheric composition is a priori unknown.
If the effective temperature is high enough that we should see Balmer lines
if the atmosphere was hydrogen-rich, then we assume a helium-rich composition.
If the temperature is too low to produce detectable Balmer lines
($T_{\rm eff} \lesssim 5000\,{\rm K}$), then the atmospheric composition
determination is based on the photometry.
Finding the atmospheric composition of a DC white dwarf solely from
  the photometry can be difficult. That being said, differences between SEDs produced by stars
  with different compositions (see Figure \ref{fig:model_grid}) can generally
  be exploited to confidently establish the atmospheric composition of such objects.

Three distinct photometric fits are performed for each cool DC white dwarf
in our sample:
one assuming a pure hydrogen
atmosphere, one assuming a pure helium atmosphere and one assuming a mixed
H/He atmosphere, where the H/He ratio is a free parameter that is adjusted
to the observations.
To identify the best of the three solutions, we do not simply pick the solution
with the smallest $\chi^2$. This approach is misguided when comparing solutions
obtained with different numbers of free parameters. In fact, it is almost
always possible to obtain a smaller $\chi^2$ by adding one more free parameter
to the fit (the H/He ratio), but the gain obtained with this additional parameter
is not always statistically significant. Instead, we
use the Akaike information criterion (AIC). This criterion penalizes a model
according to the number of free parameters it contains, thus enforcing
Occam's razor. The best model is the one with the smallest AIC, which is given
by \citep{akaike1974new,burnham2002model}
\begin{equation}
  {\rm AIC} = 2k -2 \ln L,
\end{equation}
where $k$ is the number of degrees of freedom in the model and $L$
is the maximum value of the likelihood function. Note that for least-square
model fitting the likelihood function
is simply given by $\ln L = C - \chi^2 /2$, where $C$ is a constant
that is independent from the model and can be ignored.
When we fit photometric data, we are only fitting a small number of
data points. In such circumstances, it is necessary to use an AIC that
is corrected for small data samples \citep{hurvich1989regression},
\begin{equation}
  {\rm AICc} = {\rm AIC} + \frac{2k(k+1)}{n-k-1},
\end{equation}
where $n$ is the sample size (in our case the sample size corresponds
to the number of bands included in our photometric fit).

To our knowledge, this is the first time that such a detailed statistical
analysis is applied to the problem of the determination of the
atmospheric composition of DC stars. Previous analyses either only used
mixed H/He models when CIA was obviously present in the SED
\citep{bergeron1997chemical,bergeron2001photometric,kilic2010detailed}
or directly compared mixed and pure solutions without rigorously computing
the significance of the $\chi^2$ reduction following the addition of
a third free parameter \citep{kowalski2006phd,kilic2009spitzer}.\footnote{
A third approach, proposed by \cite{saumon2014near}, consists of plotting $\chi^2$ as a
function of the H/He abundance ratio and checking if $\chi^2(\rm H/He)$
has a significant minimum. Note, however, that no quantitative criterion
was used to define what makes a $\chi^2(\rm H/He)$ minimum statistically
significant.}
Both approaches are problematic. The first one could miss mixed objects
that do not show strong CIA and the second one is biased towards mixed
solutions (e.g., $\approx 50\%$ of DCs analyzed by \citealt{kilic2009spitzer}
are classified as mixed objects). By accurately quantifying the weight
of evidence in favor of each solution, our methodology should not be
influenced by those biases. 

We performed a small numerical experiment that clearly illustrates the
danger of using an approach where the H/He ratio
is directly considered as a free parameter (i.e., by fitting every DC star
with a grid of mixed H/He compositions ranging from pure hydrogen to pure
helium and adjusting the H/He ratio as a free parameter).
We generated a set of 100 ``synthetic'' pure hydrogen stars with effective
temperatures ranging from 4000 to 6000\,K.
To do so, we used the synthetic photometry obtained from our models
to which we added a Gaussian noise to mimic the observational
uncertainties (we assumed a typical 3\% uncertainty for the visible
photometry and 5\% for the infrared).
For 45\% of these pure hydrogen objects, the fit obtained by adjusting the H/He ratio
in order to obtain the smallest $\chi^2$ corresponds to a mixed H/He solution
(i.e., $-5 < \log {\rm H/He} < 0.5$).\footnote{Note that pure helium models are
  less susceptible to this bias, as the same experiment with pure helium models
  shows that 10\% are (incorrectly) classified as mixed objects.} This numerical experiment
demonstrates that this approach is heavily biased toward finding mixed solutions, even for
stars that actually have a pure composition. In contrast, if we use our AIC approach, we find
the correct (pure hydrogen) solution in all cases.

We also performed an experiment to make sure that our AIC approach is not
  biased toward pure compositions. To do so, we fitted a set of 100 ``synthetic''
  mixed H/He white dwarfs with effective temperatures ranging from 4000 to
  6000\,K and hydrogen abundances between $\log {\rm H/He}=-5$ and $0$.
  As in our previous experiment, we injected a Gaussian noise to mimic the
  observational uncertainties. We find that our AIC
  approach retrieves a mixed solution for 92\% of these objects.\footnote{While we
    find a mixed solution in 92\% of cases, we find the correct H/He ratio
    in 68\% of cases. This difference is due to the degeneracy
    between the SEDs of mixed objects with a high H/He ratio and a low H/He ratio
    (see Section~3.1 of Paper~II).}
  The cases where our AIC approach fails to retrieve a mixed solutions
  are all for objects with hydrogen abundances near $\log {\rm H/He}=0$,
  where the SED becomes very similar to that of a pure hydrogen object.
  Therefore, the bias of the AIC approach against
  mixed objects is very weak compared to the bias against pure compositions
  exhibited by the method where the H/He ratio is directly considered as a free parameter,
  which justifies our methodology.

The AIC can also be used to estimate how likely it is that our
atmospheric composition determinations are accurate \citep{burnham2002model}.
For each model $i$ (i.e., for the pure hydrogen, the pure helium and the mixed H/He models)
we can compute the difference in AIC with respect to the best model,
\begin{equation}
  \Delta_i ({\rm AIC}) = {\rm AIC}_i - {\rm min}\,{\rm AIC}.
\end{equation}
From there, the Akaike weights $w_i$ are computed as,
\begin{equation}
  w_i = \frac{\exp \left[ -\frac{1}{2} \Delta_i ({\rm AIC}) \right]}
  {\sum_j \exp \left[ -\frac{1}{2} \Delta_j ({\rm AIC}) \right]}.
\end{equation}
Each weight $w_i$ can be interpreted as the probability that model $i$
is the best model. Therefore, we can now quantify the degree of
confidence of our atmospheric composition determinations. In particular,
given our hydrogen-rich and helium-rich fits, we can compute the odds
that a DC star is hydrogen-rich versus helium-rich. Some examples of
the usefulness of the Akaike weights are given in Section~\ref{sec:DADC_fit}.

\subsubsection{DQs and DZs}
For metal-polluted objects, we still rely on a photometric fit (see
Section~\ref{sec:DA}) to determine $T_{\rm eff}$ and $\log g$, but we
also use the observed spectrum to find the abundances of heavy elements.
Our approach is identical to that of
\cite{dufour2005detailed,dufour2007spectral}. Once a photometric solution
is found, we adjust the metal abundances to obtain a good fit to the
spectroscopy. As the abundances derived from this spectroscopic fit are
usually different from our initial guess, we repeat the whole procedure
(including the photometric fit) until a consistent solution is found.
For DQ and DQpec objects, we use the C$_2$ Swan bands to constrain the C/He
abundance ratio. For DZs, we use the \ion{Ca}{2} H \& K doublet and
the \ion{Ca}{1} resonance line to find Ca/He. The Fe/He, Mg/He and Na/He
ratios are also adjusted when the relevant spectral lines are visible.
Otherwise, the abundance ratios of all heavy elements are scaled to the abundance
of Ca to match the abundance ratios of CI chondrites \citep{lodders2003solar}.\footnote{There is an exception to this rule. For Fe, we set the default
abundance to half that predicted by the chondritic Fe/Ca number ratio.
This is a rough way of taking the faster diffusion of Fe 
\citep{paquette1986diffusion,koester2009accretion,hollands2017cool}
into account when Fe lines are not detected.}
Finally, unless there is direct evidence of the presence of hydrogen in the
atmosphere (i.e., Balmer lines, CH bands or H$_2-$He CIA), a hydrogen-free
atmosphere is assumed for all DQ and DZ white dwarfs.

\section{Results}
\label{sec:results}
Following the fitting procedures described in Section~\ref{sec:fitting_procedure}, we obtained
the atmospheric parameters of all 501 objects included in our sample. They are listed in
Table~\ref{tab:sol}, which is published in its entirety in the machine-readable format.
Our statistical analysis of this sample and its implication on the spectral evolution of 
cool white dwarfs is presented in Section~\ref{sec:evolution}. The rest of this section
is devoted to a more in-depth analysis of a number of noteworthy objects.

\begin{deluxetable*}{lccccccc}
  \tabletypesize{\footnotesize}
  \tablecaption{Atmospheric parameters \label{tab:sol}}
  \tablehead{PSO & $T_{\rm eff}\,$(K)
    & $\log g$ & $M\,(M_{\odot})$  & $\tau_{\rm cool}\,$(Gyr)\tablenotemark{a}
    & $\log\,{\rm H/He}$ &
    $\log\,{\rm C/He}$ & $\log\,{\rm Ca/He}$}
    \startdata
    J000224.474+635745.512   &  4565(50) &  7.758(0.024) &  0.432(0.013) &  4.45 &     He & --    & --     \\
    J000410.478$-$034008.751 &  6955(60) &  7.931(0.023) &  0.551(0.013) &  1.40 &      H & --    & --     \\
    J000720.831+123018.402   &  4885(45) &  8.090(0.021) &  0.625(0.014) &  6.45 &     He & --    & --     \\
    J000728.917+340342.227   &  5545(50) &  8.153(0.021) &  0.666(0.013) &  5.02 &     He & --    & --     \\
    J000754.487+394731.294   &  4680(25) &  6.768(0.039) &  0.081(0.016) &  1.40 &      H & --    & --     \\
    J000935.139+310840.332   &  7960(290)&  8.053(0.100) &  0.609(0.062) &  1.21 &     He & --    & --     \\
    J001122.399+424038.082   &  6990(50) &  7.975(0.017) &  0.577(0.010) &  1.47 &      H & --    & --     \\
    J001214.188+502514.381   &  6445(40) &  8.247(0.012) &  0.746(0.008) &  3.24 &      H & --    & --     \\
    J001412.352$-$131109.057 &  5855(35) &  8.217(0.016) &  0.724(0.010) &  4.10 &      H & --    & --     \\
    J001737.914$-$051650.624 &  5630(40) &  7.918(0.026) &  0.537(0.015) &  2.40 &      H & --    & --     
 \enddata
 \tablenotetext{a}{Cooling times are computed using the evolutionary models of \cite{fontaine2001potential}.}
 \tablecomments{Table~\ref{tab:sol} is published in its entirety in the machine-readable format.}
\end{deluxetable*}

\subsection{DAs and DCs}
\label{sec:DADC_fit}
\begin{figure*}
    \includegraphics[width=\linewidth]{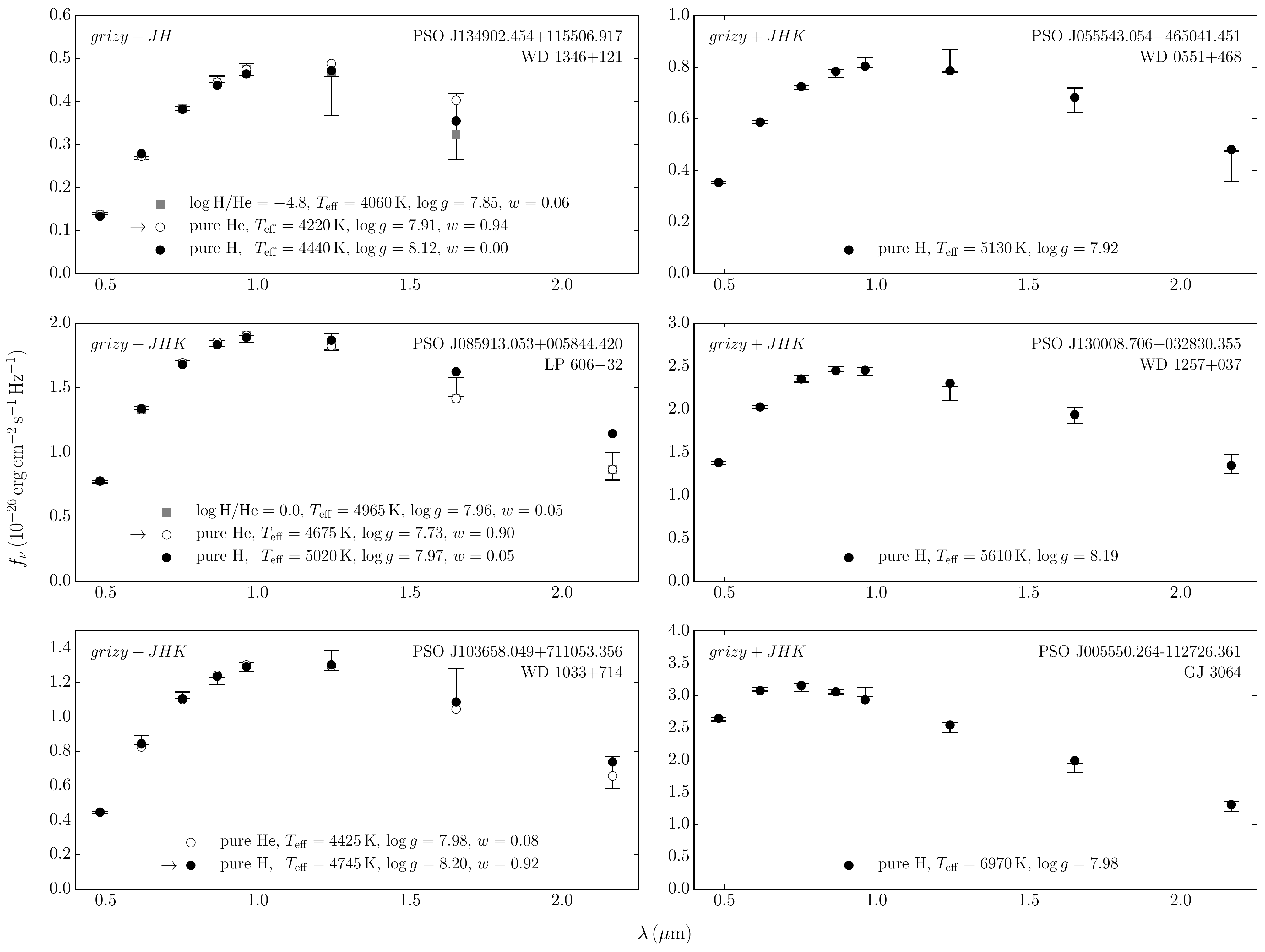}
    \caption{Examples of photometric fits to DC (left column) and DA white dwarfs (right column).
    For DCs, pure hydrogen, pure helium and mixed solutions are shown (the best fit is indicated
    by the small arrow next to the legend).
    Note that the mixed H/He solution is only shown if it corresponds
    to a different solution than the pure hydrogen and pure helium solutions.
    The complete set of photometric fits (459 objects) is available in the
    online journal.}
  \label{fig:DADC_photometry}
\end{figure*}

Figure~\ref{fig:DADC_photometry} displays some examples of photometric fits to DA and DC white dwarfs (the complete set of photometric fits is available in the online journal).
For DCs, both pure hydrogen and pure helium solutions are shown. The mixed H/He solution is only shown
if the solution found when the H/He ratio was adjusted as a free parameter does not correspond to
a pure hydrogen or a pure helium solution.
For each object, the composition that leads to the smallest Akaike weight is designated as being 
the most likely composition (Section~\ref{sec:procedure_DC}). 

\begin{figure}
    \includegraphics[width=\columnwidth]{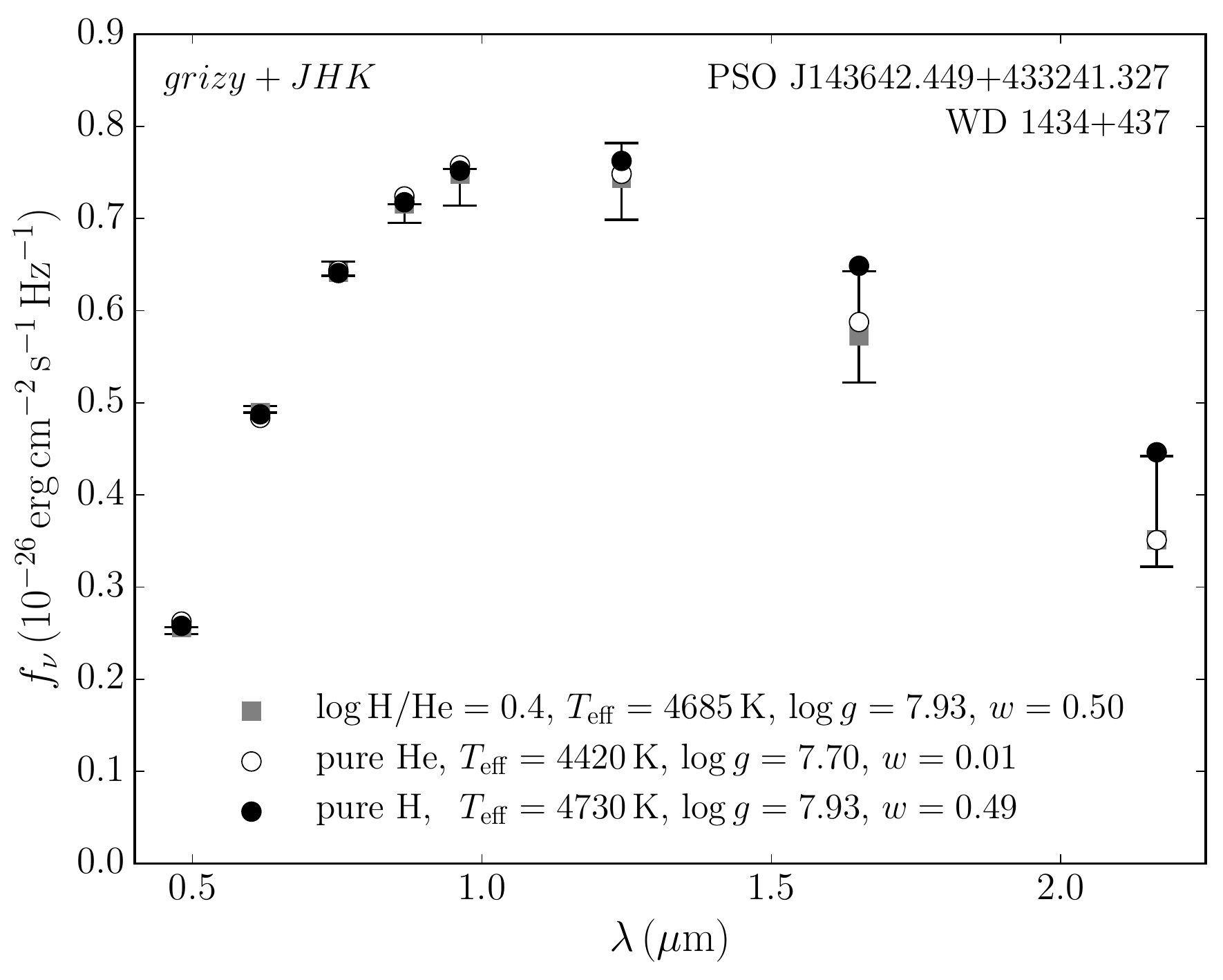}
    \caption{Photometric fits to the DC white dwarf WD 1434+437. It is hard to tell the atmospheric
      composition of this object.}
  \label{fig:akaike_ex}
\end{figure}

Note that the Akaike weights are particularly useful when trying to assess our degree of
confidence in the determination of the atmospheric composition of DC stars.
For instance, for a star like WD 1033+714 (Figure~\ref{fig:DADC_photometry}), the
Akaike weights are $w_{\rm H}=0.93$ for the pure hydrogen solution, 
$w_{\rm He}=0.07$ for the pure helium solution and $w_{\rm mixed}=0.00$ for the mixed H/He solution.
This indicates that we can be quite confident that it is a hydrogen-rich object.
Similarly, the Akaike weights strongly suggest that WD~1346+121 has a pure helium atmosphere
($w_{\rm He}=0.94$, see Figure~\ref{fig:DADC_photometry}). Note that if we had simply used the
$\chi^2$ to identify the atmospheric composition of WD~1346+121, the mixed model would
have been chosen since its $\chi^2$ is smaller than that of the pure hydrogen and pure helium
solutions. By adding a penalty for the additional free parameter, the Akaike weights reverse this verdict
and assign a low probability to the mixed solution.
Figure~\ref{fig:akaike_ex} shows an example where large observational errors lead to a much more
uncertain composition. The Akaike weights for the mixed and pure hydrogen solutions are virtually
equal, which reflects our inability of confidently identifying the atmospheric composition of this object.

Another interesting aspect of Figure~\ref{fig:DADC_photometry} is the presence of two 
$T_{\rm eff}<5000\,{\rm K}$ DC white dwarfs for which we find a pure helium solution.
Note that the Akaike weights for the pure helium solutions are $\geq 0.9$ in both cases,
which suggests that we can be quite confident in those solutions.
In fact, out of the 74 DC stars cooler than 5000\,K in our sample, we find pure helium
solutions for 16 of them. We also find that 3 of those 74 DC stars have a mixed H/He
atmosphere dominated by helium, which implies that $\approx 25\%$ of DCs cooler than
5000\,K have helium-rich atmospheres.
This result is in strong disagreement with the conclusions of 
\cite{kowalski2006phd}, \cite{kowalski2006found} and \cite{kilic2009spitzer,kilic2009near},
according to which helium-rich DC white dwarfs cooler than 5000\,K are nearly nonexistent.
The implications of this finding on the spectral evolution of cool white dwarfs 
will be discussed in Section~\ref{sec:evolution}.

\subsubsection{Peculiar non-DAs}
In the $6200\,{\rm K} \leq T_{\rm eff} \leq 7600\,{\rm K}$ temperature range,
\cite{bergeron1997chemical,bergeron2001photometric} have identified DC stars
for which the photometric fit yields a pure hydrogen solution. Given their surface temperatures,
those objects are expected to show H$\alpha$ in their spectra if they have a hydrogen-rich
atmosphere. The pure hydrogen photometric fits are therefore in contradiction with the spectroscopic
observations, which led \cite{bergeron1997chemical,bergeron2001photometric} to designate those
stars as peculiar non-DAs. Furthermore, they interpreted these peculiar non-DAs as forming a distinct
physical group and, because of their proximity to the blue edge of the non-DA gap, they
speculated that they are objects whose atmospheres are about to become hydrogen-rich.

Our analysis also reveals the presence of 13 objects for which the pure hydrogen composition
inferred from the photometry is incompatible with the absence of H$\alpha$ (see Table~\ref{tab:pecNonDA}).\footnote{Note that in all 13 cases the best mixed solution corresponds either to a pure hydrogen composition
or to a mixed H/He composition that should give rise to a detectable H$\alpha$ feature.}
Interestingly, out of the seven peculiar 
non-DAs in \cite{bergeron1997chemical,bergeron2001photometric}
that are part of our sample, only two (PSO J091554.370+532508.067 and PSO J104153.917+141545.969)
are found to have an SED that is better represented by a hydrogen-rich atmosphere. In principle,
if peculiar non-DAs did form a physical group, we should find that
objects identified as peculiar non-DAs by \cite{bergeron1997chemical,bergeron2001photometric}
are also flagged as peculiar non-DAs in our analysis. These conflicting results could be
due to the fact that we rely on different photometric systems. In particular, we note that our 2MASS
infrared photometry is of lesser quality than the $JHK$ photometry used in 
\cite{bergeron1997chemical,bergeron2001photometric}. To evaluate the impact of the photometric system,
we performed photometric fits for the nine peculiar non-DAs identified in 
\cite{bergeron1997chemical,bergeron2001photometric} using their photometric data, but our atmosphere
model grid. We found that the photometric fits favor a pure hydrogen composition for five of those objects
and a pure helium composition for the remaining four. Note that we reach the same conclusions if we
use the trigonometric parallaxes reported in \cite{bergeron1997chemical,bergeron2001photometric} instead
of the more accurate \textit{Gaia} parallaxes.

\begin{deluxetable}{llc}
  \tabletypesize{\footnotesize}
  \tablecaption{Peculiar non-DAs in our sample \label{tab:pecNonDA}}
  \tablehead{PSO & MWDD object ID & $T_{\rm eff}\,$(K)}
  \startdata
  J021148.605+711911.081  &                 NLTT 7194 & 4790 \\
  J023759.111+163809.334  &               LP 410$-$67 & 5250 \\
  J095120.241+190009.463  &             WD 0948+192 & 5460 \\
  J084901.422+443934.262  & 2MASS J08490170+4439355 & 5810 \\
  J234612.394+115849.053  &          PM J23462+1158 & 5935 \\
  J081411.267+484526.630  &             WD 0810+489 & 6515 \\
  J142747.971+053230.563  &             WD 1425+057 & 6725 \\
  J141143.197+220644.803  &             WD 1409+223 & 6785 \\
  J104718.292+000717.423  &             WD 1044+003 & 6965 \\
  J104153.917+141545.969  &               V* CY Leo & 7140 \\
  J105734.525$-$073122.157  &                 LAWD 34 & 7155 \\
  J091554.370+532508.067  &                EGGR 250 & 7170 \\
  J122619.639+183634.295  &             WD 1223+188 & 7465 
  \enddata
\end{deluxetable}

The fact that the choice of the photometric system and small differences in model grids can
tip the photometric solution to another atmospheric composition raises some doubt about
the interpretation according to which peculiar non-DAs from a distinct physical group.
Could the existence of those objects simply be explained by random errors in the observations?
After all, the difference between a pure hydrogen and a pure helium solution is not always 
statistically significant (i.e., the Akaike weights can be close to each other), 
which inevitably leads to a number of misclassifications.
Given the Akaike weights of our best solution for every DC
with $T_{\rm eff}>5000\,{\rm K}$ (that is, objects for which we should see H$\alpha$ if they
had a hydrogen-rich atmosphere), we can actually compute how many peculiar non-DAs we expect due to
the uncertainties of our photometric fits. For the 45 DCs in this temperature range, we
find an average Akaike weight of 0.76 for the best solutions. This means that we expect
that the composition inferred from the photometry will be incorrect for $(24 \pm 6)\%$ of 
those 45 DCs, which is perfectly consistent with our actual error rate
in this temperature range of $27\%$ 
(12 out of 45).\footnote{For comparison,
  \cite{bergeron2001photometric} have a $33\%$ error rate for DCs between 5000\,K
  and 8300\,K (i.e., the maximum surface temperature of stars included in our sample).}

\begin{figure*}
    \includegraphics[width=\linewidth]{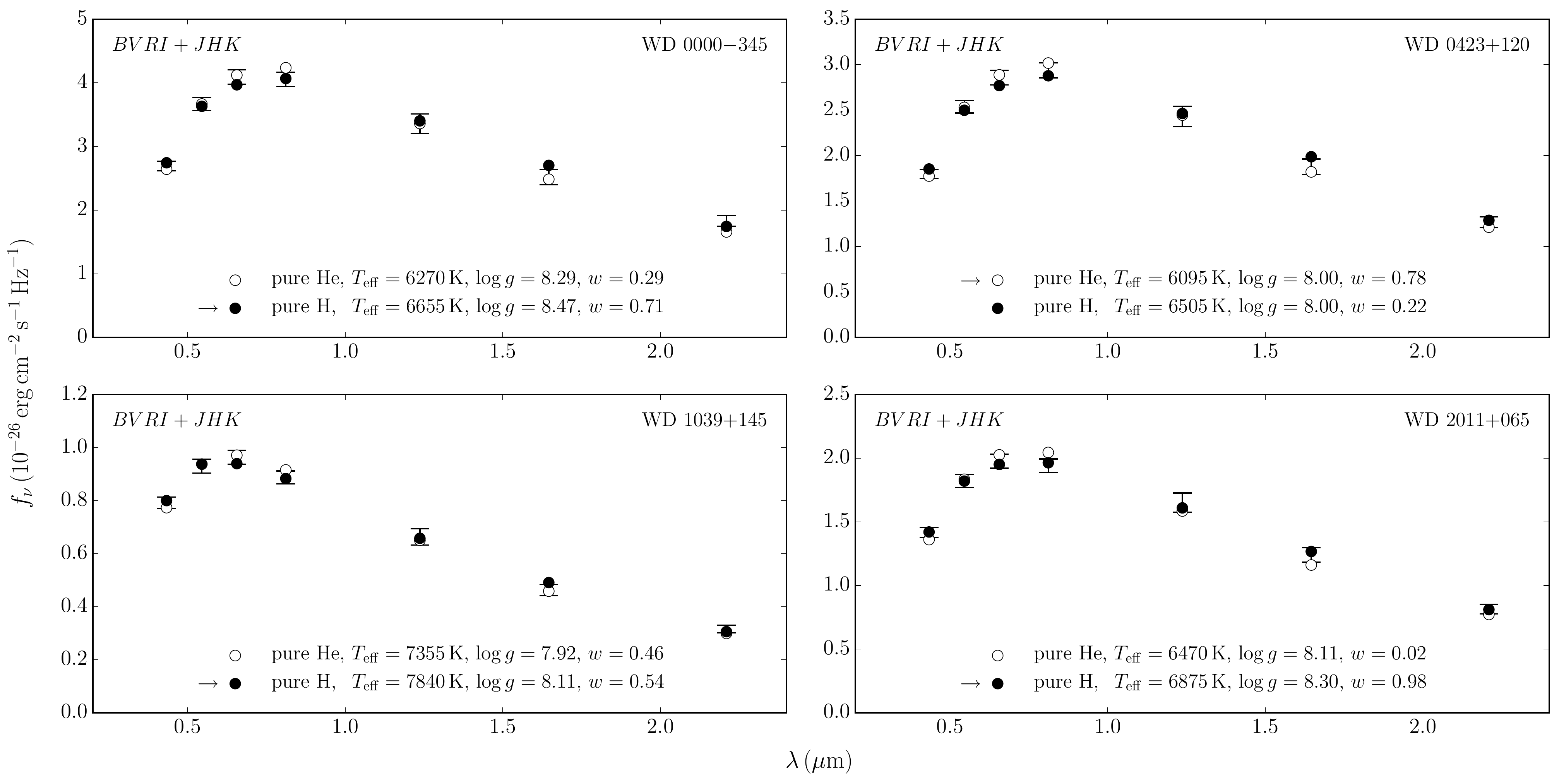}
    \caption{Photometric fits to four peculiar non-DAs identified in \cite{bergeron1997chemical,bergeron2001photometric}.
    These fits were obtained using their photometric data and trigonometric parallaxes (as they did not have any
    parallax measurement for WD~0423+120, we assumed $\log g=8$ for this object).}
  \label{fig:peculiar}
\end{figure*}

While the number of peculiar non-DAs in our sample seems consistent with random errors,
we note that \cite{bergeron1997chemical,bergeron2001photometric} identified systematic
trends in the observed SEDs of peculiar non-DAs that supported the
interpretation according to which they form a distinct physical group. In particular, they
found that the pure helium models systematically fail to reproduce the observed $B$ and $I$
photometry by underestimating the $B$ flux and overestimating the $I$ flux. The four clearest
examples of this behavior are given in Figure~17 of \cite{bergeron1997chemical} 
and Figure~11 of \cite{bergeron2001photometric}. In Figure~\ref{fig:peculiar}, we show our fits to those
four objects. While we do observe that our pure helium models slightly underestimate the flux in the
$B$ band and overestimate the flux in the $I$ band, the differences are smaller than 
$\approx 1 \sigma$ (except for WD~2011+065) and might not be significant.
Moreover, for WD~0000$-$345 and WD~1039+145,
the Akaike weights of the pure helium solution are too high to confidently rule out this solution, and, 
in the case of WD~0423+120, we even find that
the pure helium solution is a better fit to the data than the pure hydrogen solution.

As an additional test, we fitted the nine peculiar non-DAs of \cite{bergeron1997chemical,bergeron2001photometric}---all
stars for which they found a better match to pure hydrogen than to pure helium models---using
their observations and their atmosphere models. For the pure helium solutions, we find an
average Akaike weight of 0.44, meaning that the quality of the pure hydrogen and pure helium solutions is
very similar and that the chemical composition determination is quite uncertain.
Moreover, for three of those objects, we actually find a smaller $\chi^2$ (and a larger Akaike weight)
for the pure helium solution than for the pure hydrogen solution, suggesting that they are not peculiar non-DAs
after all.

Our analysis makes a strong case for a simple interpretation according to which the existence
of peculiar non-DAs is the unavoidable consequence of the intrinsic uncertainty of the photometric technique.
That being said, the fact that two objects---PSO~J091554.370+532508.067 (WD~0912+536) and
PSO~J104153.917+141545.969 (WD~1039+145)---are
flagged as peculiar non-DAs in both our analysis and the analysis of \cite{bergeron1997chemical,bergeron2001photometric}
prevents us from ruling out the idea that {\it some} peculiar non-DAs are part of a distinct physical group.
Given that the same conclusion is reached using both different models and different observations,
it appears that those two objects are indeed peculiar. Interestingly, WD~0912+536
harbors a very strong magnetic field \citep[$B \sim 100$\,MG,][]{angel1972new,angel1978magnetic},
possibly hinting to a connection between peculiar non-DAs and magnetism
\citep[however, no circular polarization was detected for WD~1039+145,][]{angel1981magnetic}.
In the same vein, we note that
\cite{bergeron1997chemical} have suggested the existence of a relation between magnetic
fields and the chemical evolution of cool white dwarfs.

\subsubsection{Hydrogen traces in helium-rich DC white dwarfs}
\label{sec:DC_H_traces}
Based on a comprehensive analysis of white dwarfs discovered by the \textit{Gaia} mission,
Bergeron et al. (2019, submitted) recently claimed that
pure helium white dwarfs below $T_{\rm eff}=11\,000\,{\rm K}$ are extremely rare.
In a nutshell, additional electron donors (metals or hydrogen) have to be included in
the atmosphere models of helium-rich objects in order to obtain reasonable masses.
Formally, only upper limits on the hydrogen abundance of DC stars can be determined.
As the presence of hydrogen traces affects the atmospheric parameters obtained from
a photometric fit, our inability to accurately measure the
H/He abundance ratio of those objects implies that our $T_{\rm eff}$ and $\log g$
determinations are uncertain.

\begin{figure}
    \includegraphics[width=\columnwidth]{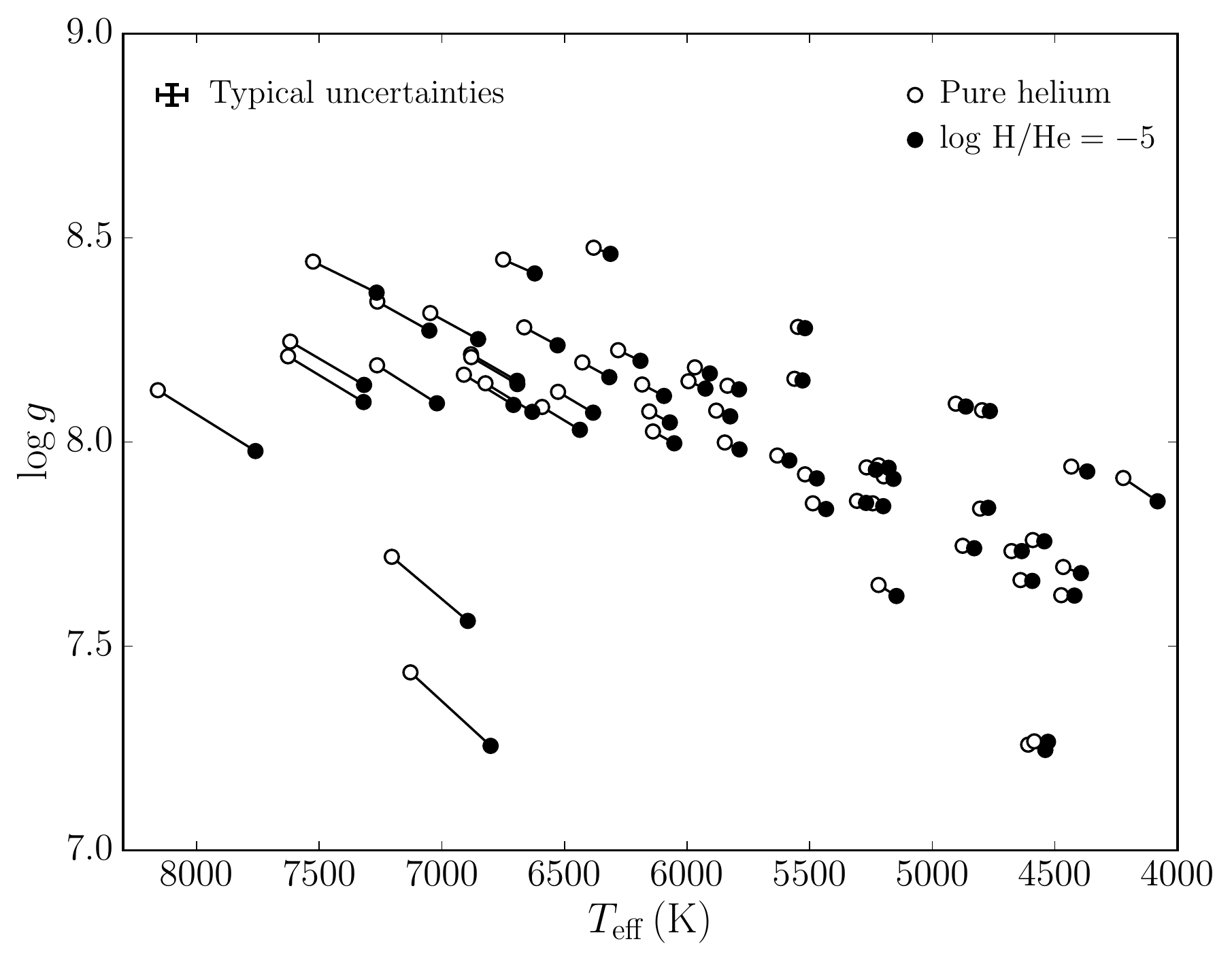}
    \caption{Surface gravity of helium-rich DC white dwarfs in our sample as a
      function of their effective temperatures. The open circles correspond to
      our pure helium solutions and the filled circles represent the solutions
      found if $\log\,{\rm H/He}=-5$ is assumed.
      Typical uncertainties on $\log g$ and $T_{\rm eff}$ are shown in the top-left corner.}
  \label{fig:DC_H5}
\end{figure}

To see by how much $T_{\rm eff}$ and $\log g$ can be affected by trace amounts of
hydrogen, we studied a subsample made up of all DC stars for which we found a pure
helium composition. For each of those objects, we performed a second photometric
fit assuming a hydrogen abundance of $\log\,{\rm H/He}=-5$. Note that this
hydrogen abundance is too small to produce any detectable H$\alpha$ feature or
to give rise to significant H$_2-$He CIA.
As shown in Figure~\ref{fig:DC_H5}, the addition of this trace amount of hydrogen
has an important effect on $T_{\rm eff}$ and $\log g$, particularly at high
temperatures. This implies significant uncertainties on the atmospheric parameters of
helium-rich DC white dwarfs.
By adding free electrons to the atmosphere, small amounts
of hydrogen reduce the determined effective temperature (by 400\,K at worst and by 120\,K
on average) and the surface gravity (to keep the luminosity constant, the star has to
be inflated to compensate the decrease of $T_{\rm eff}$). This behavior is analogous
to the reduction of the photometric $T_{\rm eff}$ and $\log g$ following the addition
of carbon---another electron donor---in
DQ model atmospheres \citep[see Figure~8 of][]{dufour2005detailed}.
To account for this uncertainty, we shifted our helium-rich solutions for DC
white dwarfs halfway between the pure helium and the $\log\,{\rm H/He}=-5$ solutions
and we adjusted the uncertainties so that they now encompass both solutions.
The $T_{\rm eff}$, $\log g$, $M$ and $\tau_{\rm cool}$ values reported in Table~\ref{tab:sol}
include these corrections.

\subsection{DZs}
\begin{figure*}
    \includegraphics[width=\linewidth]{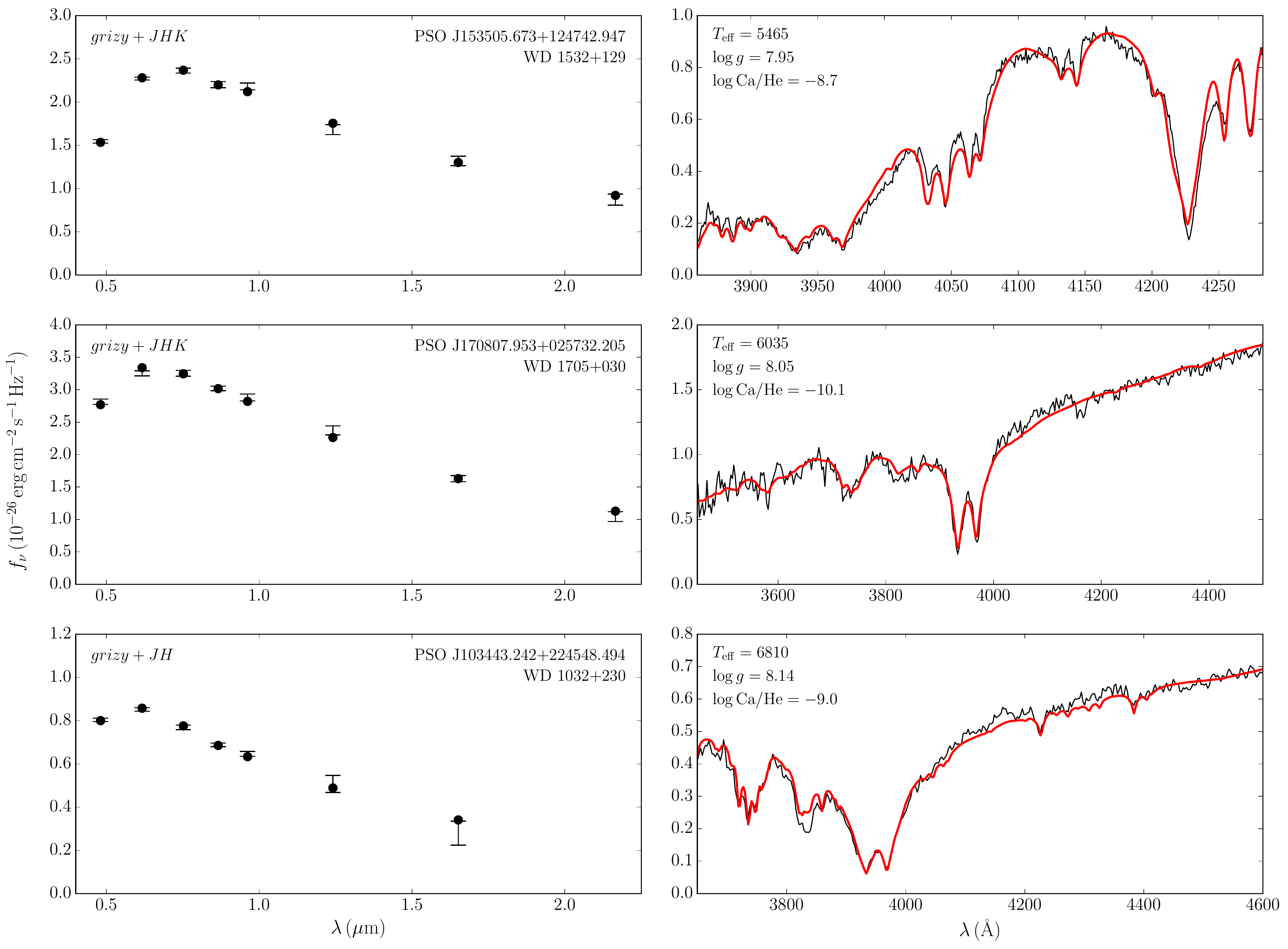}
    \caption{Examples of photometric and spectroscopic fits for three DZ white dwarfs. Each row
      represents one object. For the spectroscopic fits, the observations are shown in black
      and the best fit is represented by the red spectrum. The complete set of fits (16 objects) is
      available in the online journal.}
    \label{fig:DZ_example}
\end{figure*}

Figure~\ref{fig:DZ_example} shows examples of photometric and spectroscopic fits
to DZ white dwarfs.
The complete set of fits is available in the online journal, except for
LP~658$-$2, Ross~640, SDSS J080440.63+223948.6 and WD~2251$-$070
(PSO J055509.987$-$041037.276, PSO J162824.563+364623.925, PSO J080440.637+223945.828
and PSO J225355.149$-$064701.663), for which
we directly use the values reported in Papers I and II and in
\cite{blouin2019line}.
As in Papers I, II and III, our models are in excellent agreement
with the observations. Our solutions are consistent across all wavelengths and they
properly reproduce the observed spectral lines. In particular, for the three examples
shown in Figure~\ref{fig:DZ_example}, we obtain good fits to the resonance line of
\ion{Ca}{1} at 4226\,{\AA} and to the \ion{Ca}{2} H \& K doublet. This suggests that
the pressure and temperature structure of our DZ models are accurate, since these
profiles are sensitive to the physical conditions in the line-forming regions
of the atmosphere \citep{allard2014caii}.

\subsection{DQs}
\begin{figure*}
    \includegraphics[width=\linewidth]{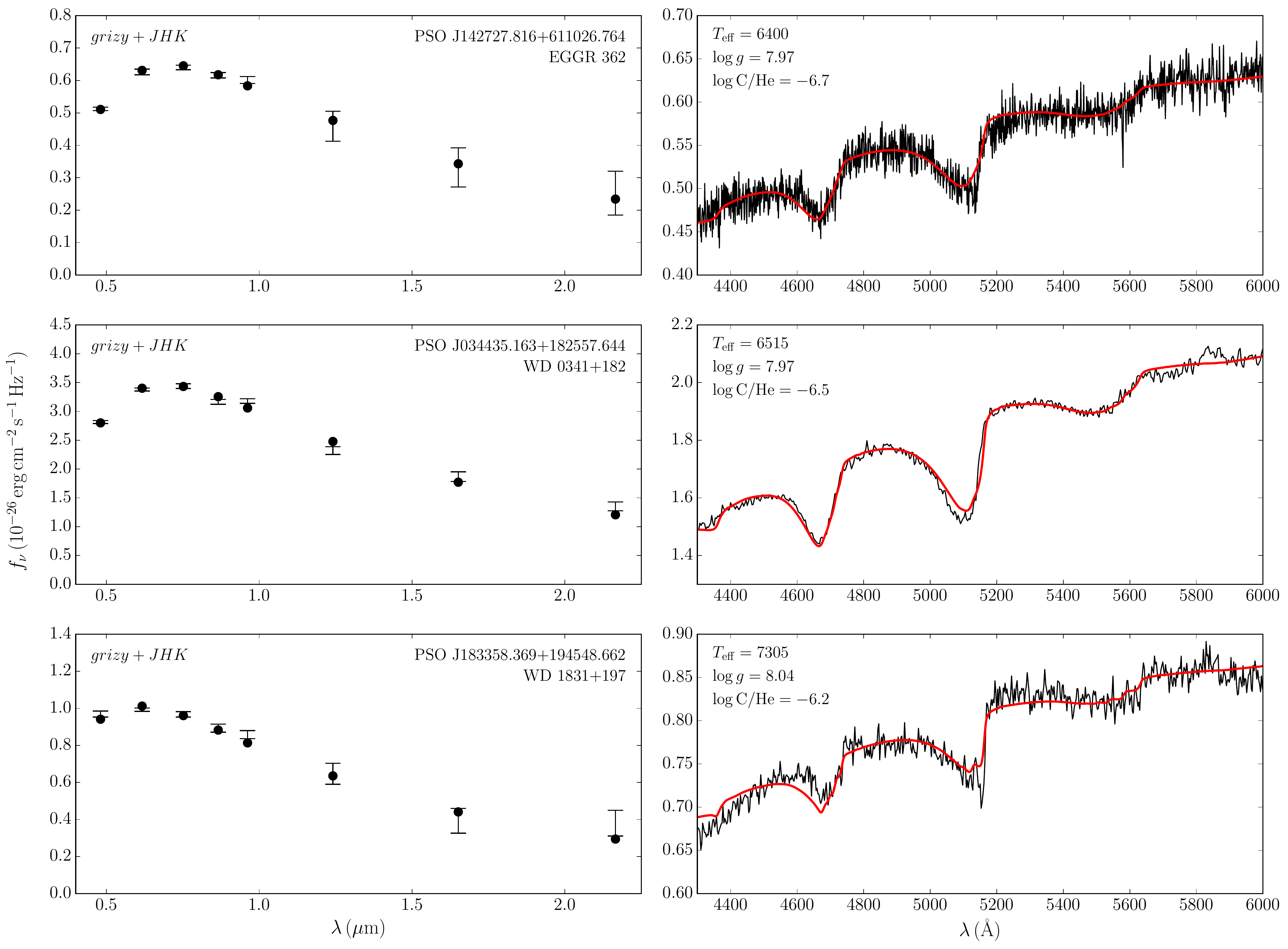}
    \caption{Similar to Figure~\ref{fig:DZ_example}, but for DQ/DQpec white dwarfs. Note that
      a shift parameter of $\alpha=0.2$ was assumed for the C$_2$ Swan bands
      (see Section~\ref{sec:alpha}). The complete set of fits (22 objects) is
      available in the online journal.}
  \label{fig:DQ_example}
\end{figure*}

In Figure~\ref{fig:DQ_example}, we show examples of photometric and spectroscopic fits
to three DQs (the complete set of fits is available in the online journal).
Clearly, our best solutions are in good agreement with the
observations. Things get more complicated if we look at cooler carbon-polluted
atmospheres. The cooling sequence of DQs stops at $T_{\rm eff}\approx 6000\,{\rm K}$
\citep{dufour2005detailed,koester2006new} to give way to DQpec white dwarfs. 
The spectra of those objects are vaguely similar to that of DQs: they show Swan-like
absorption bands that are blue-shifted by a few hundreds Angstroms
\citep{bergeron1994peculiar,schmidt1995nature,hall2008c2}. Using density functional
theory (DFT) calculations, \cite{kowalski2010origin} has convincingly shown that the physical
explanation for the DQ$\rightarrow$DQpec transition is the pressure-driven distortion
of the C$_2$ Swan bands. As a carbon-polluted white dwarf cools down, its atmosphere
becomes denser. Near $T_{\rm eff}=6000\,{\rm K}$, the helium density becomes high
enough to affect the electronic levels of the C$_2$ molecule, which leads to
an increase of the electronic transition energy $T_e$ of the Swan bands. It is this
increase that explains the shift of the C$_2$ Swan bands towards lower wavelengths.
According to the ab initio calculations of \cite{kowalski2010origin}, the shift of $T_e$ is
linear with respect to density,
\begin{equation}
\Delta T_e \,(\rm{eV}) \approx \alpha \,\rho \,(\rm{g\,cm}^{-1}),
\label{eq:dqpec}
\end{equation}
with $\alpha=1.6$.
However, as pointed out by \cite{kowalski2010origin}, such a shift overestimates the
distortion of Swan bands in the spectra of DQpec white dwarfs. There are two ways to
solve this problem: either we change the slope of Equation \ref{eq:dqpec} or we
find a way of decreasing the photospheric density of DQpec models.

\begin{figure}
    \includegraphics[width=\linewidth]{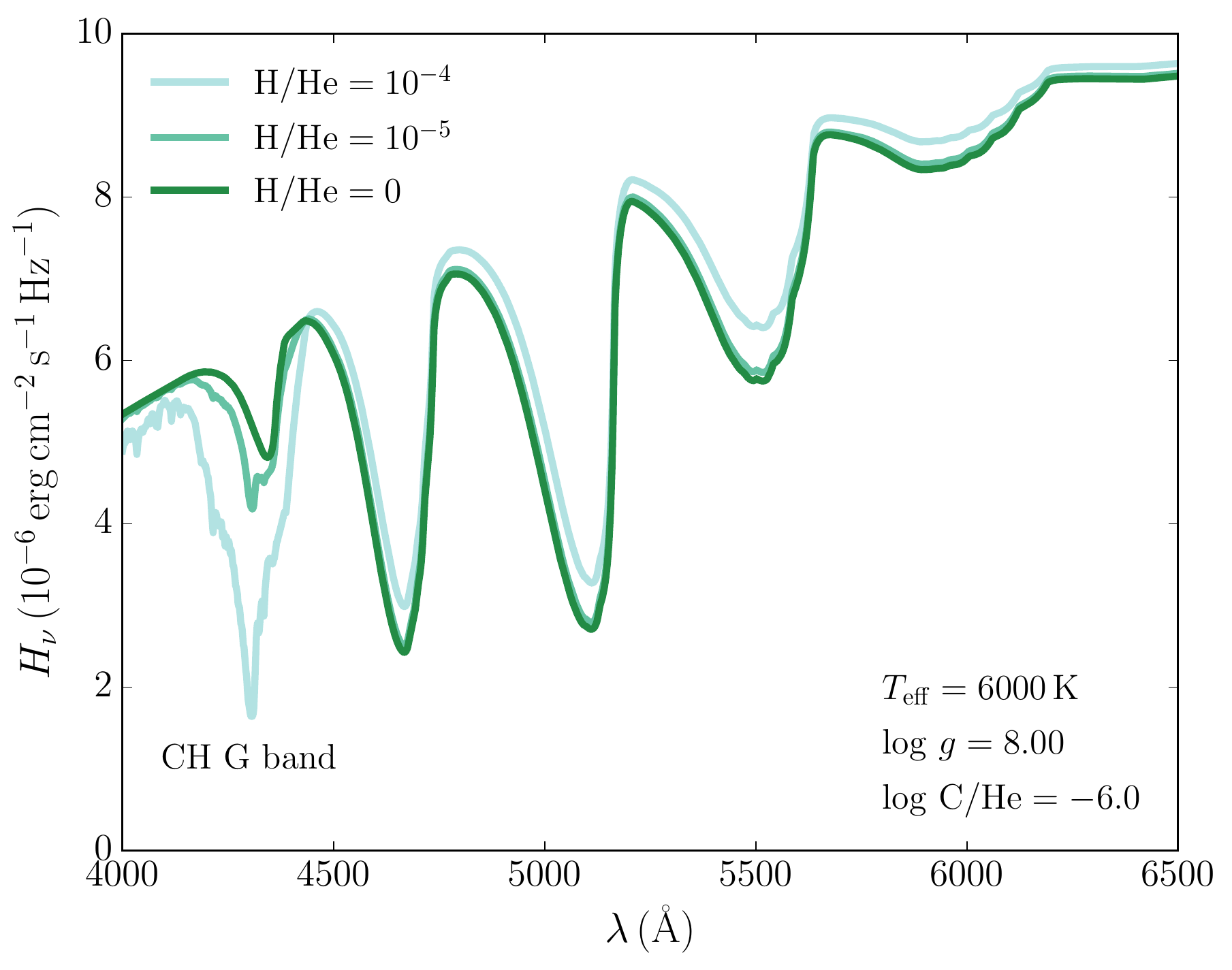}
    \caption{Synthetic spectra computed with different H/He abundance ratios and for
      atmospheric parameters similar to those of LHS 290. Note that a shift
      parameter of $\alpha=0.2$ was assumed for the C$_2$ Swan bands 
      (see Section~\ref{sec:alpha}).}
  \label{fig:ch}
\end{figure}

\cite{kowalski2010origin} proposed that the photospheric density of DQpec stars could
be lowered if hydrogen is added to their atmospheres. In fact, in their
analysis of LHS 290, they show a fit where they compensate for the strong
$\Delta T_e = 1.6 \rho$ shift by supposing an ad hoc $\log\,{\rm H/He}=-2.2$ hydrogen abundance.
We are very skeptical that this is the correct way of reconciling the DFT calculations
with the observations, since such a high amount of H would lead to the formation
of a significant quantity of CH. We added the CH rovibrational bands to our code
using the Kurucz linelists\footnote{\url{http://kurucz.harvard.edu}},
which rely on data from \cite{masseron2014ch}. Figure~\ref{fig:ch} shows synthetic
spectra computed for parameters very similar to those of LHS 290 ($T_{\rm eff}=6000\,{\rm K}$
and $\log\,{\rm C/He}=-6.0$) and with different H/He abundance ratios. Clearly, a
hydrogen abundance as low as $\log\,{\rm H/He}=-4.0$ is sufficient to produce an unmissable CH 
G band near 4300\,{\AA} and the $\log\,{\rm H/He}=-2.2$ value can be safely rejected.
For the models of Figure~\ref{fig:ch}, the pure helium and the $\log\,{\rm H/He}=-4.0$ models
have photospheric densities of $0.20\,{\rm g\,cm}^{-3}$ and $0.16\,{\rm g\,cm}^{-3}$ (at $\tau_R=2/3$), respectively.
This slight reduction of the density is not sufficient to compensate for the too important
shift implied by the $\alpha=1.6$ parameter. Therefore, we conclude that adding ad hoc
amounts of hydrogen in the atmospheres of DQpec stars is not the solution to the discrepancy
between the DFT calculations and the observations.

\subsubsection{Calibration of $\alpha$}
\label{sec:alpha}
While we rejected the idea that an undetected amount of hydrogen could bring the density
down to a level where the $\alpha=1.6$ value is compatible with the observations, it is
still possible that a missing piece of input physics in our models leads to an overestimation
of the photospheric density in cool carbon-polluted white dwarfs. However, we believe that it is more
likely that the current implementation of the shift of C$_2$ Swan bands is incorrect.
In particular, we still use the standard Swan band spectrum, which we merely shift at
every atmospheric layer with the $\Delta T_e$ value given by Equation \ref{eq:dqpec}.
Until the absorption of C$_2$ in dense helium is modeled, which is a much more
computationally intensive task than computing $\Delta T_e (\rho)$,
this approximation will remain unjustified.

\begin{figure*}
    \includegraphics[width=\linewidth]{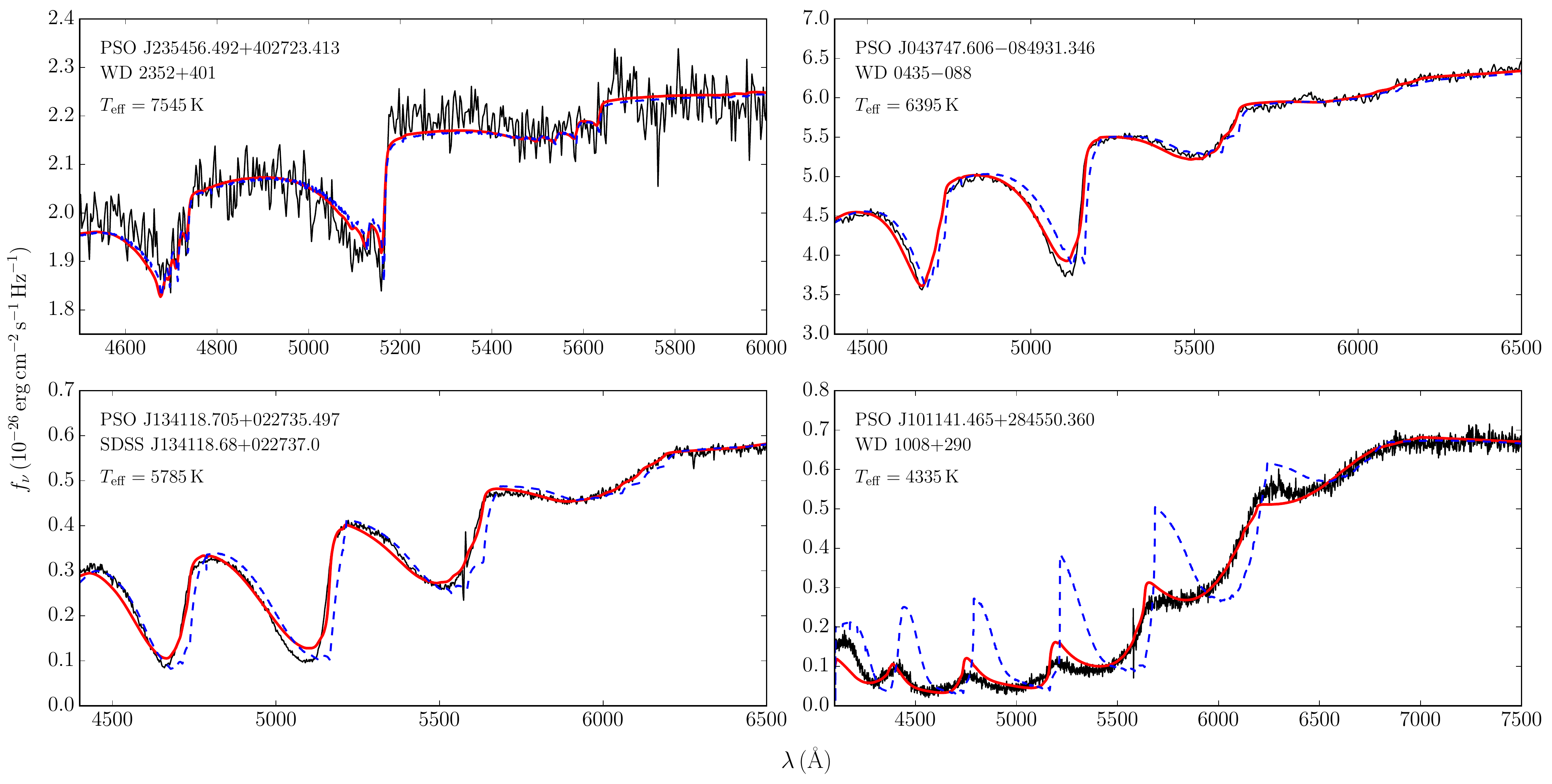}
    \caption{Spectroscopic fits to the Swan bands of four DQ/DQpec stars. These four
    stars represent a sequence of decreasing temperature, which also corresponds to
    a sequence of increasing photospheric density and of increasing distortion of
    the Swan bands by high-pressure effects. The red spectra (solid lines) correspond to our
    best solutions assuming $\alpha=0.2$ and the blue ones (dashed lines) correspond to the
    case where the Swan bands are not distorted by high-pressure effects (i.e., $\alpha=0$).
    Note that the blue spectra were computed assuming the same atmospheric parameters
    as those found by fitting the observational data with a grid of $\alpha=0.2$ models.}
  \label{fig:dqpec}
\end{figure*}

In the meantime, for the purpose of this paper, it is sufficient
to resort to a more approximative way of fitting DQpec objects
and we calibrate $\alpha$ using the observed spectra of distorted Swan bands. After
testing several $\alpha$ parameters, we found that the value that leads---for most objects---to the
best agreement between models and observations is $\alpha=0.2$.
This value is actually a compromise, since we found that no fixed $\alpha$
parameter could perfectly reproduce the distorted Swan bands of all DQpec
white dwarfs. Nevertheless, as shown in Figure~\ref{fig:dqpec}, the
$\alpha=0.2$ value allows a good fit to the Swan bands of carbon-polluted
white dwarfs across a large temperature range.
The most important disagreement is for WD~1008+290, where the bands could
be more rounded. Note, however, that additional mechanisms might play a role,
since WD~1008+290 has an intense
magnetic field \citep[$B > 10\,{\rm MG}$,][]{schmidt1999discovery}
that could affect the position of the Swan bands
\citep{liebert1978magnetic,bues1991preliminary,bues1999final}.

\subsubsection{Problematic objects}
In our analysis of DQ and DQpec objects, we encountered five cases for which a
completely satisfying solution could not be found. Here, we detail the challenges
posed by those objects and speculate about possible ways of
resolving the discrepancies between models and observations.

\begin{figure*}
    \includegraphics[width=\linewidth]{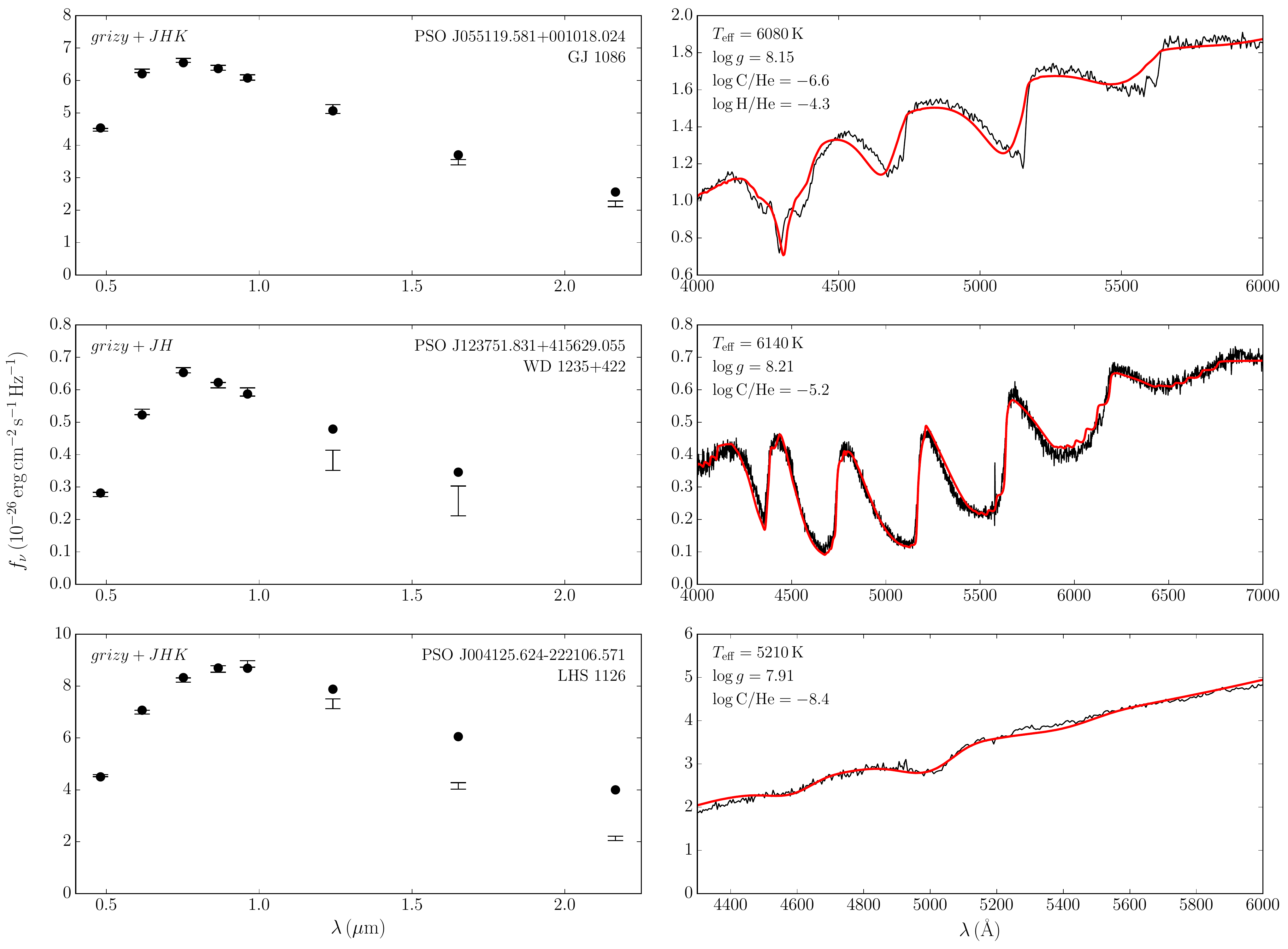}
    \caption{Similar to Figure~\ref{fig:DQ_example}, but for three
      problematic carbon-polluted white dwarfs. Note that our fits
      of WD~1235+422 and LHS~1126 are based on hydrogen-free models
      (see text for details).}
  \label{fig:dq_weird}
\end{figure*}

\paragraph{GJ 1086} GJ~1086 (G~99$-$37) is one of the only two known white dwarfs
to show a CH G band (the other one being BPM~27606). Using a hydrogen abundance
of $\log\,{\rm H/He}=-4.3$, we are able to achieve a reasonable fit to the G band
(Figure~\ref{fig:dq_weird}). However, our solution overestimates the distortion of
the Swan bands (the C$_2$ bands of our fit are too rounded compared to the observations).
This problem suggests that GJ~1086 contains more hydrogen than assumed, since adding hydrogen 
would reduce the distortion of the Swan bands by decreasing the photospheric density.
However, the H/He ratio is already constrained by the strength of the CH G band,
and thus we cannot find any solution that simultaneously fits all spectral features.
One way to fix this problem would be to change the C$_2$ and/or CH dissociation equilibrium. After all,
pressure effects are known to affect the H/H$_2$ ratio in dense hydrogen
\citep{vorberger2007hydrogen,holst2008thermophysical} and in dense helium-rich 
mediums \citep{kowalski2006dissociation}, so it is likely that something similar occurs with
C$_2$ and CH in the dense atmospheres of cool carbon-polluted white dwarfs.
If the nonideal effects on the dissociation 
equilibria are such that they reduce the CH/C$_2$ ratio, they could explain (at least in part)
the discrepancy described above. Furthermore, it is also worth noting that GJ~1086 is a
magnetic white dwarf \citep{angel1974determination}, with $B \approx 7\,{\rm MG}$ 
\citep{berdyugina2007molecular,vornanen2010gj}. As already stated, the impact of such strong
fields on the opacities of carbon-polluted atmospheres remains unclear.
Note also that convection might be suppressed \citep{tremblay2015evolution,gentile2018magnetic},
which would significantly affect the atmosphere structure.

\paragraph{WD 1235+422} WD~1235+422 is characterized by strong Swan bands that
are very well represented by our models (Figure~\ref{fig:dq_weird}).
However, our fit to the near-infrared photometry is less satisfying.
While our overestimation of the flux in the $J$ and $H$ bands might be due to
observational errors, it is also possible that we are missing an absorption source in
the infrared. Given the effective temperature of WD~1235+422, H$_2-$He CIA cannot explain
this flux depletion. In Section~\ref{sec:c2hecia},
we explore the possibility that the infrared flux depletion of WD~1235+422 is
due to C$_2-$He CIA. It is also worth noting that WD~1235+422 is another
magnetic white dwarf \citep{vornanen2013spectropolarimetric}.

\paragraph{LHS 1126} \cite{wickramasinghe1982infrared} were the first to identify
the strong near-infrared flux deficit that characterizes the SED of LHS~1126 (GJ~2012).
\cite{bergeron1994peculiar} and \cite{giammichele2012know} have explained this flux deficit
as being due to H$_2-$He CIA (they find mixed H/He compositions of 
$\log\,{\rm H/He}=0.1$ and $-1.2$, respectively).
However, their solutions are not compatible with the ultraviolet
observations of LHS~1126, since \cite{wolff2002element} found that Ly$\alpha$ is
better reproduced with $\log\,{\rm H/He}=-5.5$. Another puzzling finding is that the
near and mid-infrared energy distribution of LHS~1126 fits a Rayleigh-Jeans spectrum
mimicking a $T_{\rm eff}>10^5\,{\rm K}$ blackbody \citep{kilic2006mystery}.
As with GJ~1086 and WD~1235+422, we also have to worry about
the usual suspect---magnetism---since
spectropolarimetric measurements of LHS~1126 cannot rule out the presence of
a $B<3\,{\rm MG}$ magnetic field \citep{schmidt1995nature}.

As all previous analyses of LHS~1126, our fit is far from satisfactory. We interpreted the
small depression at $\approx 5000\,${\AA} as being due to distorted Swan bands and, based on the
depth of this feature, we concluded that $\log\,{\rm C/He}=-8.4$. Regarding our fit to
the photometry, we find a hydrogen abundance of $\log\,{\rm H/He}=-1.2$ if
we adjust the H/He ratio to the near-infrared photometry.
As explained above, this solution must be rejected because 
of the Ly$\alpha$ analysis of \cite{wolff2002element}. The origin of the infrared flux depletion 
remains unknown and for this reason the solution displayed in Figure~\ref{fig:dq_weird}
was obtained assuming a hydrogen-free atmosphere and ignoring the $JHK$ bands
in our fit to the photometric data.
Based on novel ab initio calculations, \cite{kowalski2014infrared} proposed that
He$-$He$-$He CIA might explain the infrared flux depletion of LHS~1126. This opacity source
is included in our models, but fails to explain the SED of LHS~1126. As noted by
\cite{kowalski2014infrared}, this failure may be due to our poor constraints 
on the ionization equilibrium of helium under high-pressure conditions 
\citep{fortov2003pressure,kowalski2007equation}, which controls the photospheric 
density and therefore the intensity of He$-$He$-$He CIA. That being said, our sample
contains many helium-rich objects that have denser photospheres than LHS~1126
(e.g., the 16 DC stars cooler than 5000\,K for which we find a pure helium composition)
and none of them shows a discrepancy similar to that observed here.

\paragraph{WD 1036$-$204 and WD 1008+290} Those two objects
---identified in Table \ref{tab:sol} as PSO J103855.315$-$204049.761 and
PSO J101141.465+284550.360---present a very similar problem.
For both of them, our models were unable to
properly match the photometric observations. In particular, we were forced to ignore
the $g$ and $r$ bands in order to obtain an adequate fit to the other photometric bands.
We do not know what explains this discrepancy, but we note that both objects have a very
low effective temperature ($4530 \pm 215\,$K for WD~1036$-$204 and $4335 \pm 165\,$K for WD~1008+290)
and strong magnetic fields \citep{schmidt1999discovery,jordan2002search} that might affect
their structures and opacities.

\subsubsection{C$_2-$He CIA}
\label{sec:c2hecia}
For some DQs in our sample, our fits to the infrared photometry are unsatisfactory
(see for instance WD~1235+422 in Figure~\ref{fig:dq_weird}).
A possible explanation for this mismatch is the omission of an important opacity
source in our atmosphere models. Here, we investigate the possibility that C$_2-$He CIA
affects the SED of cool carbon-polluted atmospheres.
A priori, this hypothesis might seem unlikely since the peak of the C$_2-$He CIA
spectrum is expected to be in the mid infrared, beyond the $JHK$ bandpasses considered here.
In fact, the fundamental vibrational band of C$_2$ is located at $
1855\,{\rm cm}^{-1}$ \citep{herzberg1950molecular},
which implies that absorption will be especially important near 5.4\,$\mu$m. In contrast,
H$_2-$He CIA, which dominates the SED of the coolest mixed H/He white dwarfs 
\citep{bergeron2002model,kilic2012old,gianninas2015ultracool},
peaks near 2.3\,$\mu$m where the fundamental vibrational band of H$_2$ is located.
Nevertheless, it is possible that C$_2$ overtone bands, which are located at lower
wavelengths, are important enough to affect the near-infrared SED of some cool DQpec stars.

To investigate this issue, we use
the methodology presented in \cite{blouin2017pressure} for H$_2-$He CIA. More
precisely, we use ab initio molecular dynamics (MD) to simulate
the evolution of a C$_2$ molecule in a dense helium medium.
In this framework, atoms move according to classical dynamics and 
DFT is used at each time step to compute the
electronic charge density. As shown in \cite{blouin2017pressure}, this
methodology is accurate at both low and high densities, where many-body
collisions become important.

The simulations were performed with the CPMD\footnote{\url{http://cpmd.org}}
plane-wave DFT code \citep{marx2000ab,cpmd}.
The DFT calculations were carried out
using the PBE exchange-correlation functional \citep{perdew1996generalized}
and ultrasoft pseudopotentials \citep{vanderbilt1990soft}.
Each simulation consisted of one C$_2$ molecule surrounded by 15, 31, or 63 He
atoms in a cubic box whose length was adjusted to obtain the desired
density. The density-temperature space was explored with 35 distinct simulations,
ranging from $T=4000$ to 8000\,K and from $\rho=0.08$ to 1.4\,g\,cm$^{-3}$.
For each simulation, we computed at every time step the dipole moment resulting from
the total electronic charge density and the distribution of all nuclei
\citep{silvestrelli1998maximally,berghold2000general}.
At the end of the MD simulations, we obtained
the absorption spectra $\alpha(\omega)$ using the Fourier transform of
the dipole moment time autocorrelation function
\citep{silvestrelli1997ab,kowalski2014infrared,blouin2017pressure}.

We took precautions to make sure that
the periodic boundary conditions of the simulation cell do not give rise
to any artificial absorption features. To do so, we performed simulations
with different box sizes while keeping the density constant by adjusting
the number of helium atoms in each simulation. We found that absorption spectra
obtained from MD simulations performed in a box of at least
10\,au (5.3\,{\AA}) are virtually identical to those obtained from simulations
performed in larger boxes. Therefore, finite-size effects are negligible
as long as the simulation cell is at least 10\,au large and all simulations
presented here satisfy this criterion.
Note that we also verified that our simulations have converged with
respect to the MD simulation time by comparing absorption spectra obtained
for different trajectory lengths. We found that a 32\,ps trajectory is
usually sufficient to obtain a satisfactory convergence.

\begin{figure}
    \includegraphics[width=\linewidth]{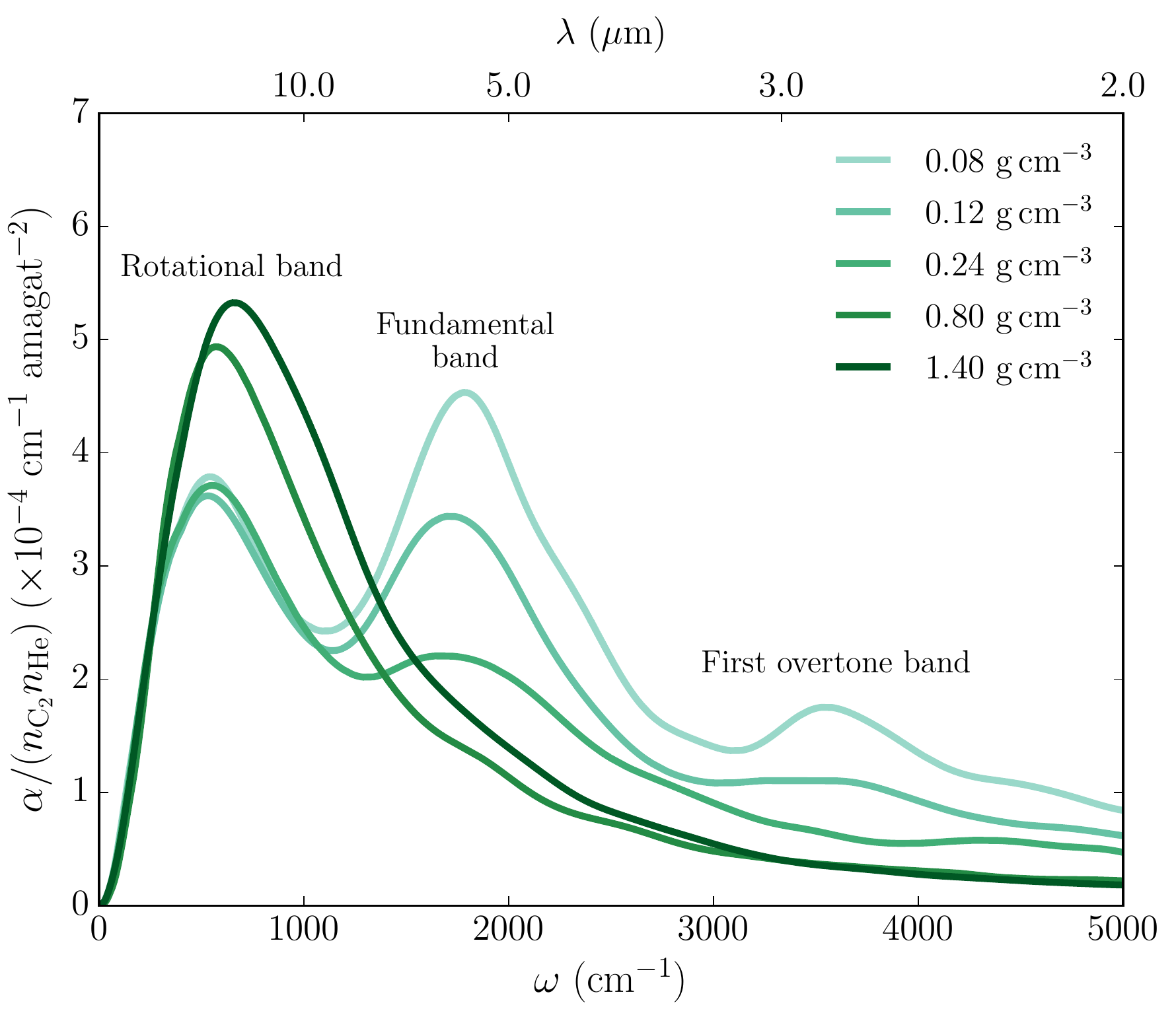}
    \caption{C$_2-$He CIA spectra for different densities at $T=4000\,{\rm K}$.
      All spectra are divided by the number density of C$_2$ and He.}
  \label{fig:c2hecia}
\end{figure}

Figure~\ref{fig:c2hecia} shows the results of our DFT-MD calculations for
different densities representative of the photosphere of cool carbon-polluted
white dwarfs. At low densities, the general shape of the C$_2-$He CIA spectrum
is very similar to that of H$_2-$He CIA
\cite[for comparison, see Figure~3 of][]{abel2012infrared},
except that the rotational and vibrational bands are located at higher
wavelengths. This is a direct consequence of the higher mass of the C$_2$ molecule.
Also similar to H$_2-$He CIA is the evolution of the absorption spectrum with
increasing density. In particular, the fundamental band becomes less and less
important, while the rotational band becomes more prominent and shifts towards
lower wavelengths \citep{blouin2017pressure}.
For the purpose of this paper, the main result of our simulations is that the 
overtone bands of the C$_2-$He CIA are too weak to significantly affect the near-infrared flux
of cool carbon-polluted white dwarfs. In fact, we implemented
C$_2-$He CIA in our model atmosphere code---using analytical fits to our results---and
we found that it only has an effect for cool ($T_{\rm eff} \leq 5000\,{\rm K}$),
relatively carbon-rich ($\log\,{\rm C/He} \geq -5$) models and that this effect is
limited to wavelengths beyond the $K$ band.

\section{The spectral evolution of cool white dwarfs}
\label{sec:evolution}
Now that we have determined the atmospheric composition of each of the 501 objects
included in our sample (Table~\ref{tab:sol}), we are ready to revisit the spectral
evolution of cool white dwarfs. Before going to the results (Section~\ref{sec:evolution_results}),
we discuss how we can account for selection biases in our sample (Section~\ref{sec:evolution_biases}).

\subsection{Correcting for biases}
\label{sec:evolution_biases}
The quantity of interest for the study of the spectral evolution of white dwarfs
is the fraction of stars that have a hydrogen-rich atmosphere in a volume complete sample,
$\rho_{\rm H} / \rho_{\rm tot} = \rho_{\rm H} / (\rho_{\rm H} + \rho_{\rm He})$.\footnote{In this work, an atmosphere with a H/He abundance ratio greater than 1 is considered hydrogen-rich.}
Unfortunately, the sample considered in this paper is not volume-complete.
We decided not to use a volume-complete sample as it would drastically reduce the number
of objects in our sample 
\citep[the almost complete 20\,pc sample contains 106 white dwarfs cooler than 8000\,K,][]{hollands2018gaia}
and significantly increase the statistical errors on $\rho_{\rm H} / \rho_{\rm tot}$.

Given the incompleteness of our sample, a correction
must be applied to relate the number of white dwarfs in our sample ($N_{\rm H}^{\rm sample}$
and $N_{\rm He}^{\rm sample}$) to the unbiased space density ratio.
Remember that the four main selection criteria (Section~\ref{sec:sample_selection}) 
are that each star must have a parallax measurement from the \textit{Gaia} DR2, 
$grizy$ photometry from Pan-STARRS1, $J$ photometry from 2MASS and a spectrum to allow 
an accurate spectral classification. By far, the two main limiting factors out of those 
four criteria are the need for $J$ photometry from 2MASS and the need for spectroscopic 
observations. Below, we examine how each of these two criteria can induce a bias on
$N_{\rm H}^{\rm sample} / N_{\rm tot}^{\rm sample}$ and how we can correct for those biases
to obtain the unbiased space density ratio $\rho_{\rm H} / \rho_{\rm tot}$.

If we ignore for the moment the criterion on spectroscopic observations, our sample is 
effectively limited by the 2MASS $J$ band, since Pan-STARRS and \textit{Gaia} have much deeper fields
than 2MASS. Depending on the effective temperature, a hydrogen-rich white dwarf can be more or 
less luminous in the $J$ band than a helium-rich object with the same effective temperature 
and surface gravity. This can induce a selection bias in favor of hydrogen-rich or helium-rich objects.
To correct for this bias, we use the $V^{\rm max}$ correction \citep{schmidt1975mass}, where $V^{\rm max}$ is
the volume defined by the maximum distance at which a given object would still appear in our sample.
More specifically, the quantity of interest is $V^{\rm max}_{\rm H} / V^{\rm max}_{\rm He}$,
the ratio between the observed volume of hydrogen-rich objects and that of helium-rich objects.
Using our model fluxes, we computed the ratio of the luminosities and, from there, we obtained
$V^{\rm max}_{\rm H} / V^{\rm max}_{\rm He}$. The corresponding $V^{\rm max}$ ratios are given
in Table~\ref{tab:vhhe}. Note that we assumed $\log g =8$ to compute $V^{\rm max}_{\rm H} / V^{\rm max}_{\rm He}$, which is justified
by the very similar mass distributions of hydrogen-rich and helium-rich white dwarfs \citep{giammichele2012know}.
The decrease of $V^{\rm max}_{\rm H} / V^{\rm max}_{\rm He}$ at low temperatures is a direct consequence of the
onset of CIA, which significantly enhances the opacity in the $J$ band for hydrogen-rich objects.
With the data of Table~\ref{tab:vhhe}, we can relate the number of hydrogen-rich and helium-rich white dwarfs in our
sample to the space density ratio,
\begin{equation}
  \frac{\rho_{\rm H}}{\rho_{\rm H} + \rho_{\rm He}} =
  \left( 1 + \frac{N_{\rm He}^{\rm sample} V^{\rm max}_{\rm H}}{N_{\rm H}^{\rm sample} V^{\rm max}_{\rm He}} \right)^{-1}.
\label{eq:correction}
\end{equation}

\begin{deluxetable}{cc}
  \tabletypesize{\footnotesize}
  \tablecaption{$V^{\rm max}$ correction for the ratio of hydrogen-rich to helium-rich objects. \label{tab:vhhe}}
  \tablehead{\colhead{\hspace{0.3cm}$T_{\rm eff}\,$(K)}\hspace{0.3cm} & 
    \colhead{\hspace{0.3cm}$V^{\rm max}_{\rm H} / V^{\rm max}_{\rm He}$}\hspace{0.3cm}}
  \startdata
  8500 & 1.15 \\
  8000 & 1.16 \\
  7500 & 1.17 \\
  7000 & 1.19 \\
  6500 & 1.21 \\
  6000 & 1.23 \\
  5500 & 1.22 \\
  5000 & 1.20 \\
  4750 & 1.18 \\
  4500 & 1.15 \\
  4250 & 1.09 \\
  4000 & 1.01 \\
  3750 & 0.82 
  \enddata
\end{deluxetable}

Let's now turn to the bias induced by the requirement of having spectroscopic observations for every
object in our sample. An appreciable fraction of our spectra (158 out of 501) are from the Sloan Digital
Sky Survey (SDSS), which is plagued by numerous selection effects. 
The SDSS consists of first imaging the sky in five bandpasses and then using this photometry to
select targets that deserve spectroscopic observations. This target selection process is based on
many different selection criteria that overlap with one another 
\citep{harris2003initial,eisenstein2006catalog}. Hence, modeling the selection biases
induced by the SDSS is an intractable task.

\begin{figure}
  \begin{center}
    \includegraphics[width=\columnwidth]{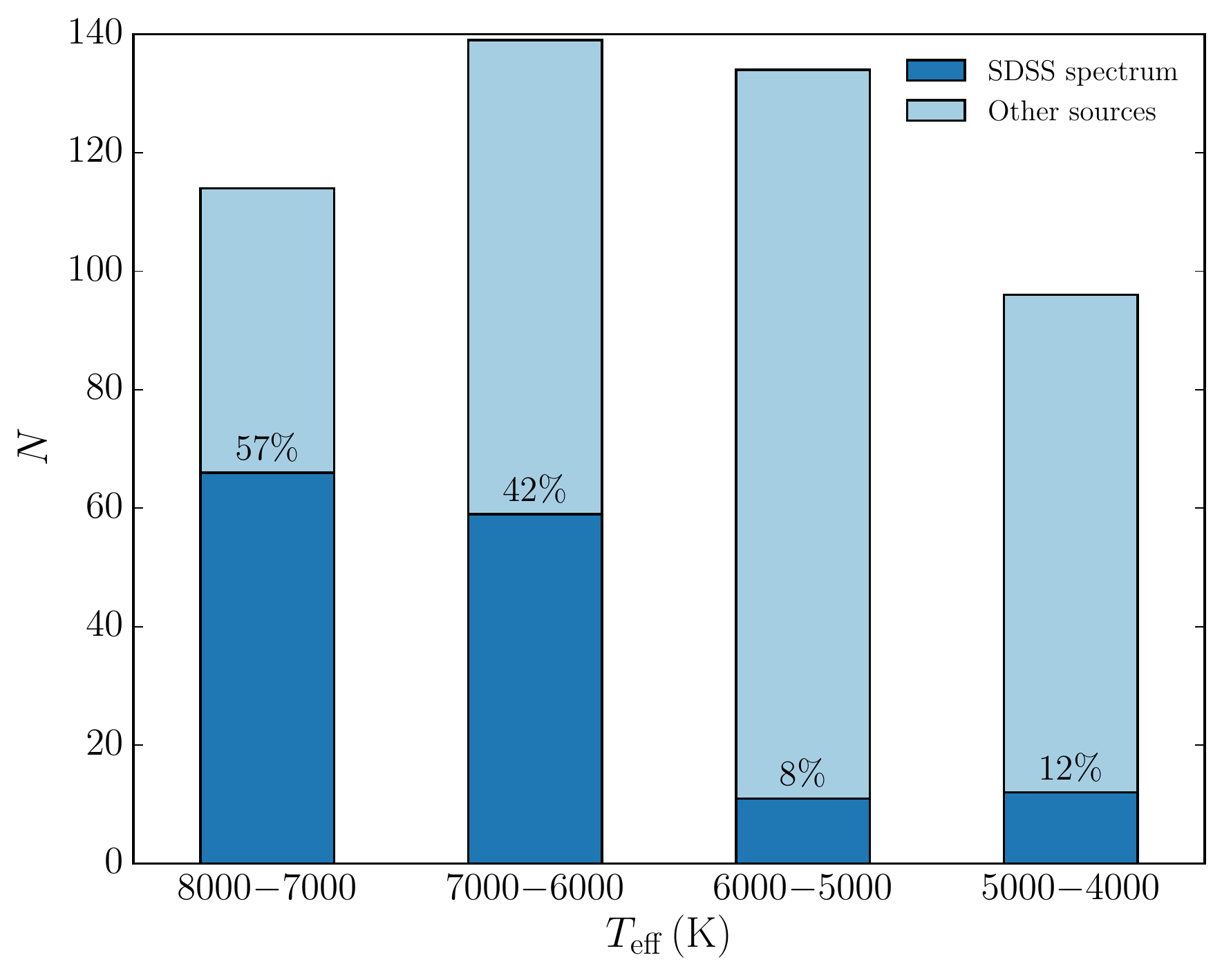}
    \caption{The total length of each bar indicates the number of objects in our sample that fall
      in each effective temperature bin. The dark blue portion of each bar represents the fraction
      of those objects that were spectroscopically observed in the SDSS.}
    \label{fig:sdss_spectra}
  \end{center}
\end{figure}

Fortunately, we do not need to worry about these selection effects for objects cooler than 6000\,K,
since for the vast majority of those cool objects we rely on spectroscopic observations from other
sources (Figure~\ref{fig:sdss_spectra}).\footnote{For the most part, objects that are not in the
  SDSS are from the \cite{limoges2015physical} sample. White dwarfs in that sample were selected
  from reduced proper motion diagrams \citep{limoges2013toward} and are therefore not subject to
  biases in favor of hydrogen-rich or helium-rich objects.}
However, above 6000\,K, SDSS objects are much more abundant and
we might be at the mercy of SDSS selection effects. 
To evaluate the severity of those selection effects, we computed the fraction of
hydrogen-rich stars for objects of our sample that were spectroscopically observed by the SDSS 
and for those that were not. We found absolutely no difference for the 6000$-$8000\,K temperature
range as a whole ($0.80 \pm 0.04$ for the SDSS objects and $0.82 \pm 0.03$ for the rest).\footnote{The uncertainties are due to the finite size of the samples and are estimated as $\sigma_f = \sqrt{f(1-f)/N}$, where $N$ is the sample size and $f$ is the fraction of objects that are hydrogen-rich. Note that $f \pm \sigma_f$ corresponds to a 68\% confidence interval.}
However, there are
differences if we look at smaller temperature bins. For the 7000$-$8000\,K bin, we found hydrogen-rich fractions
of $0.91 \pm 0.04$ for the SDSS objects and $0.84 \pm 0.05$ for the rest, and, for the 6000$-$7000\,K bin,
we found $0.68 \pm 0.06$ and $0.80 \pm 0.04$. Strictly speaking, these differences 
between SDSS and non-SDSS objects are not statistically significant. 
The probability that, given the finite size of our samples, the fraction of hydrogen-rich objects is the same
for stars observed by the SDSS and for those not observed by the SDSS (i.e., the $p$-value)
is 0.26 for the 7000$-$8000\,K bin and 0.11 for the 6000$-$7000\,K bin.
Nevertheless, it is worth keeping in mind that the SDSS might induce a small bias in favor of
hydrogen-rich objects between 7000 and 8000\,K and a small bias in favor of helium-rich
objects between 6000 and 7000\,K

\subsection{Results and discussion}
\label{sec:evolution_results}
\begin{figure}
  \begin{center}
    \includegraphics[width=\columnwidth]{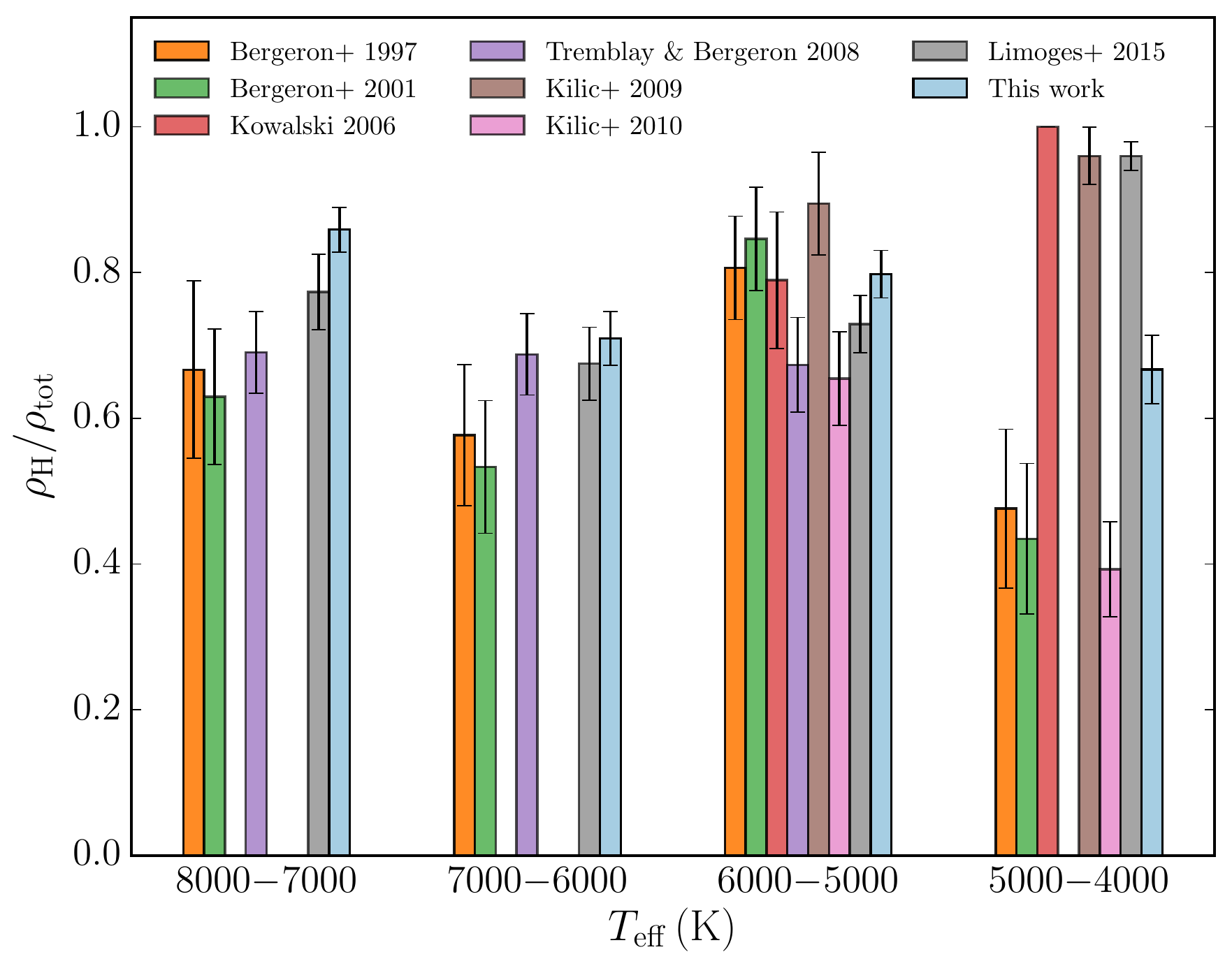}
    \caption{Fraction of hydrogen-rich white dwarfs as a function of effective temperature for
      1000\,K bins. Different colors
      represent data taken from different studies (see legend). The error bars
      indicate the $1\sigma$ uncertainty associated with the finite number of
      objects in each bin. Note that the results of our sample were corrected using Equation 
      \ref{eq:correction} and the data of Table~\ref{tab:vhhe}.}
    \label{fig:fracH_histogram}
  \end{center}
\end{figure}

Figure~\ref{fig:fracH_histogram} shows the evolution of the $\rho_{\rm H} / \rho_{\rm tot}$ ratio
as a function of decreasing effective temperature and compares our results to those of previous studies.
Down to $T_{\rm eff}=5000\,{\rm K}$, our results are in good agreement with previous studies---particularly
with \cite{limoges2015physical}, whose results were based on a bigger sample than the six other
studies shown in Figure~\ref{fig:fracH_histogram}. 
It is only below 5000\,K that our results stand out from the rest. For the 4000$-$5000\,K
temperature bin, we find $\rho_{\rm H} / \rho_{\rm tot}=0.69 \pm 0.05$, while other studies found either a
much lower \citep[$\approx 0.4$,][]{bergeron1997chemical,bergeron2001photometric,kilic2010detailed} 
or a much higher fraction \citep[$\gtrsim 0.95$,][]{kowalski2006phd,kilic2009spitzer,limoges2015physical}.

Two factors contribute to this difference. 
On the one hand, contrarily to \cite{kowalski2006phd} and \cite{kilic2009spitzer,kilic2009near},
our sample includes metal-polluted white dwarfs. As those objects are all helium-rich in the
4000$-$5000\,K temperature bin, they contribute to give rise to a lower $\rho_{\rm H} / \rho_{\rm tot}$ ratio.
On the other hand, our models (Paper I) include the nonideal
input physics \citep[in particular the improved Ly$\alpha$ opacities,][]{kowalski2006found} that led
\cite{kowalski2006phd}, \cite{kilic2009spitzer,kilic2009near} and \cite{limoges2015physical} to conclude
that the vast majority of cool DC white dwarfs are hydrogen-rich. This factor explains why we find a
higher $\rho_{\rm H} / \rho_{\rm tot}$ ratio than \cite{bergeron1997chemical,bergeron2001photometric} and
\cite{kilic2010detailed}, who did not include such high-density nonideal effects in their atmosphere models.
Note, however, that our conclusions are less extreme than \cite{kowalski2006phd}, as we find that the SEDs
of $\approx 25\%$ of DC white dwarfs cooler than 5000\,K are better fitted with helium-rich models
(Section~\ref{sec:DADC_fit}). This is the expected result. 
We know many DZ white dwarfs cooler than $T_{\rm eff}=5000\,{\rm K}$ that must have a helium-rich
atmosphere in order to explain their broad spectral lines (for examples, see \citealt{dufour2007spectral},
\citealt{hollands2017cool} and Paper III).
As the pollution of a white dwarf by rocky debris is independent from the evolution of the white
dwarf itself, a significant number of non-polluted counterparts (i.e., helium-rich DCs) must exist.

\begin{figure}
  \begin{center}
    \includegraphics[width=\columnwidth]{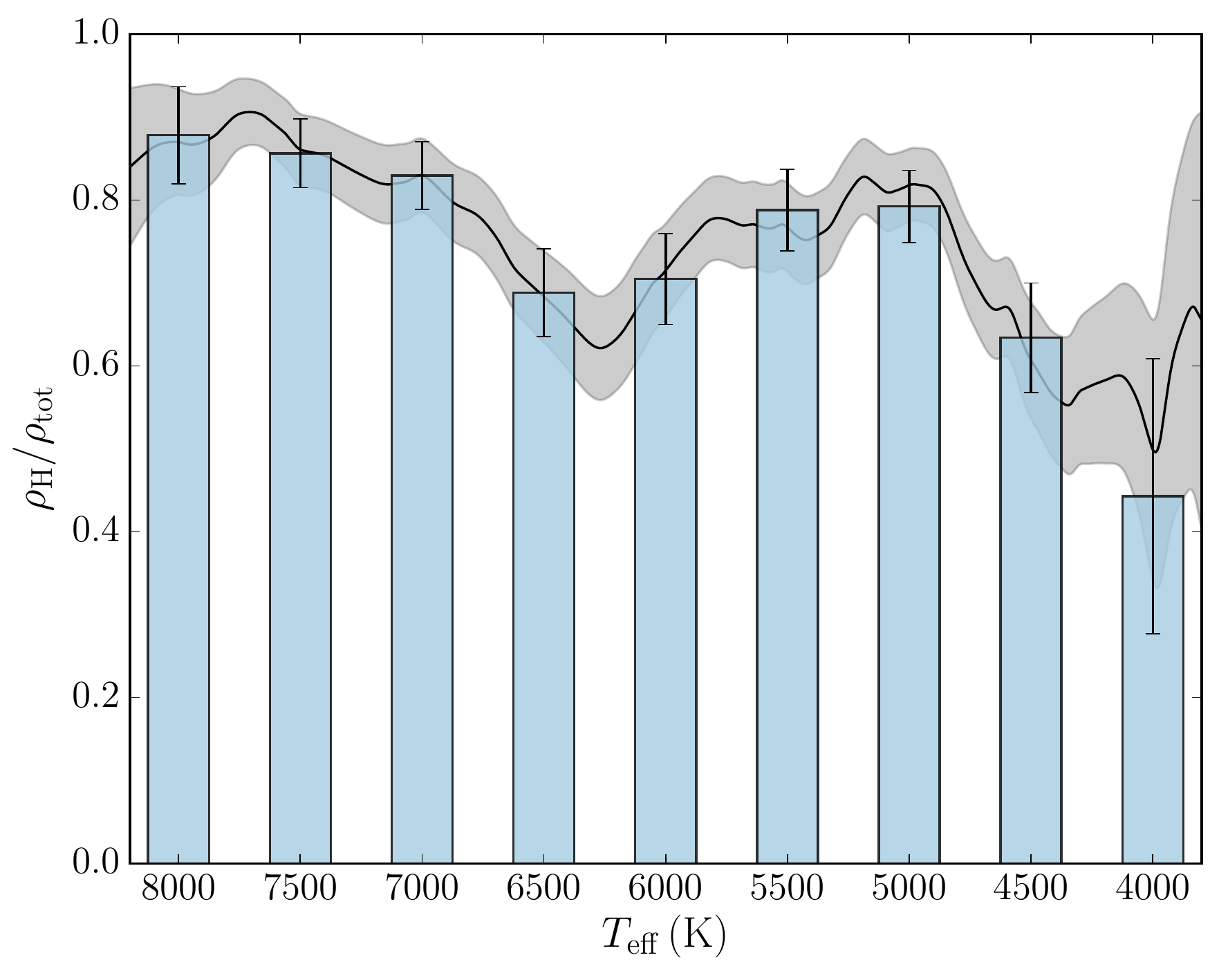}
    \caption{Fraction of hydrogen-rich white dwarfs as a function of effective
      temperature for fixed 500\,K bins (in blue) and for a 500\,K moving bin (in gray,
      see text for details).
      As in Figure~\ref{fig:fracH_histogram},
      the error bars indicate the $1\sigma$
      uncertainty associated with the finite number of objects in each bin
      and the correction given by Equation \ref{eq:correction} was applied.}
    \label{fig:fracH_histogram_500}
  \end{center}
\end{figure}

From Figure~\ref{fig:fracH_histogram}, the fraction of hydrogen-rich stars seems 
almost consistent with an evolution
at a fixed $\rho_{\rm H} / \rho_{\rm tot}$ ratio. However, the picture looks quite different if we use
smaller temperature bins. Figure~\ref{fig:fracH_histogram_500} shows the same plot as Figure~\ref{fig:fracH_histogram}, but, this time, 500\,K temperature bins were used. To make sure that no feature is missed by
our arbitrary binning choice, we also added a continuous $\rho_{\rm H} / \rho_{\rm tot}$ vs $T_{\rm eff}$ curve
to Figure~\ref{fig:fracH_histogram_500} (in gray).
This continuous curve was obtained by computing $\rho_{\rm H}/\rho_{\rm tot}$ within a 500\,K moving bin, which
eliminates the arbitrariness of the choice of the bin boundaries. Note also that the weight of each star in
the calculation of $\rho_{\rm H}/\rho_{\rm tot}$ was computed using a Gaussian with a standard deviation
given by the uncertainty on $T_{\rm eff}$.

\begin{figure}
  \begin{center}
    \includegraphics[width=\columnwidth]{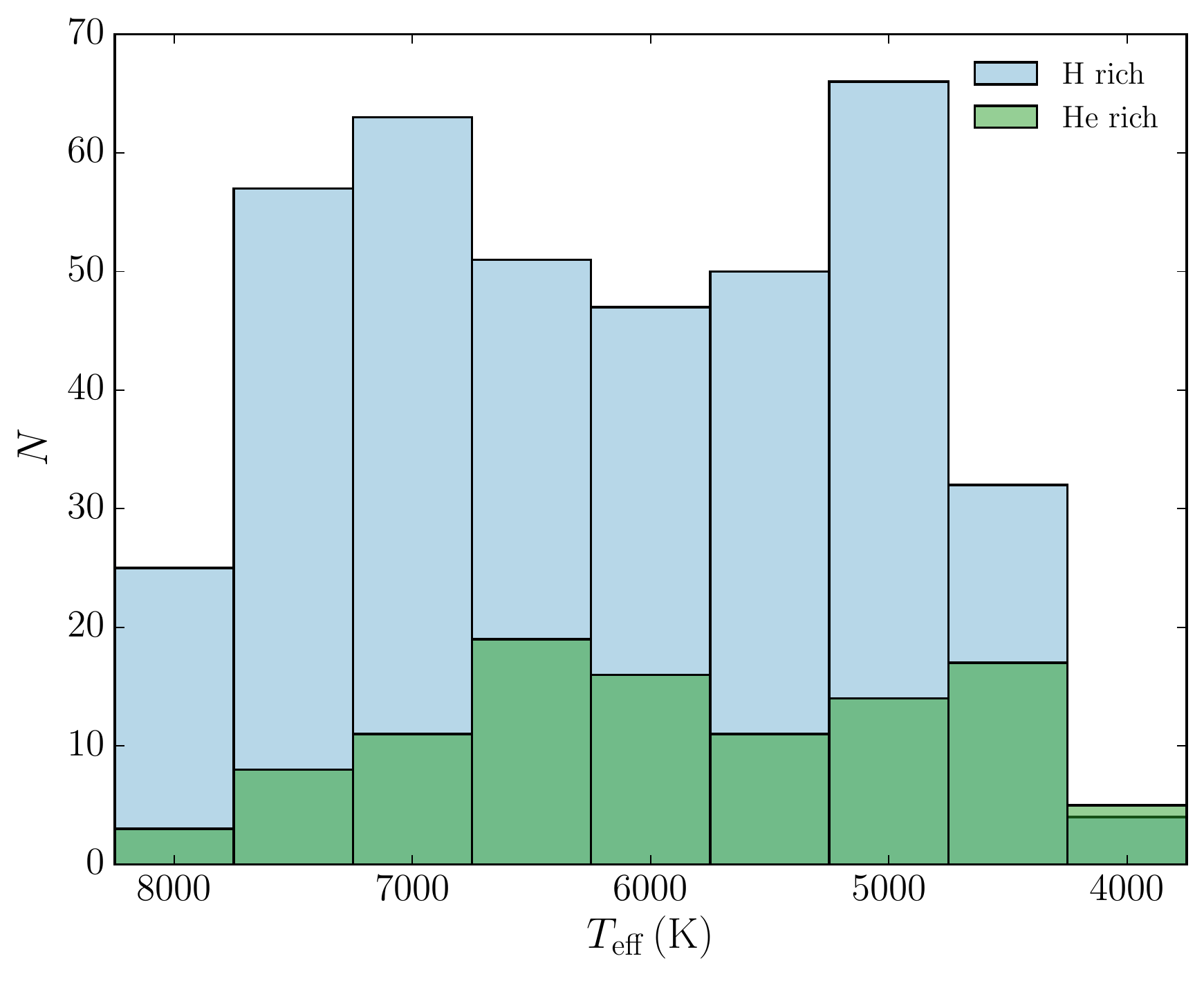}
    \caption{Effective temperature distribution of hydrogen-rich and helium-rich stars in our sample.
      An Anderson-Darling test shows that the probability that both distributions belong to the
      same parent population is $0.002$.}
    \label{fig:comp_dist}
  \end{center}
\end{figure}

Figure~\ref{fig:fracH_histogram_500} reveals a much more complex picture than Figure~\ref{fig:fracH_histogram}.
The hydrogen-rich fraction decreases from $T_{\rm eff} \approx 7500\,{\rm K}$ until
$T_{\rm eff} \approx 6250\,{\rm K}$, then increases until $T_{\rm eff} \approx 5000\,{\rm K}$, and
decreases again below 5000\,K. To confirm that these fluctuations of $\rho_{\rm H}/\rho_{\rm tot}$ are
statistically significant, we compared the effective temperature distributions of hydrogen-rich and helium-rich
stars in our sample (Figure~\ref{fig:comp_dist}).
An Anderson--Darling test shows that hydrogen-rich and helium-rich white dwarfs
do not follow the same temperature distribution, as the probability that both temperature distributions
belong to the same population is $0.002$.\footnote{The corresponding probability given by the more usual
Kolmogorov--Smirnov test is significantly higher ($0.013$). However, the Anderson--Darling test is more
appropriate, since it is more sensitive than the Kolmogorov--Smirnov test when the differences between
both distributions are more important near their extremities \citep{feigelson2012modern},
which is precisely the case here.} Moreover, note that these conclusions remain unchanged if we take
into account the $V^{\rm max}_{\rm H} / V^{\rm max}_{\rm He}$ corrections given in Table~\ref{tab:vhhe}.
If we randomly add or remove hydrogen-rich stars in each temperature bin (where the number of additions
or withdrawals is determined by the $V^{\rm max}_{\rm H} / V^{\rm max}_{\rm He}$ ratio), we find that
the probability that both the hydrogen-rich and helium-rich temperature distributions come from the same
distribution is $\approx 0.004$. Therefore, it is highly unlikely that the large increases and decreases
of $\rho_{\rm H}/\rho_{\rm tot}$ with respect to $T_{\rm eff}$ (Figure~\ref{fig:fracH_histogram_500})
are due to random fluctuations associated with small-number statistics.

\subsubsection{Behavior above 5000\,K}
\begin{figure}
  \begin{center}
    \includegraphics[width=\columnwidth]{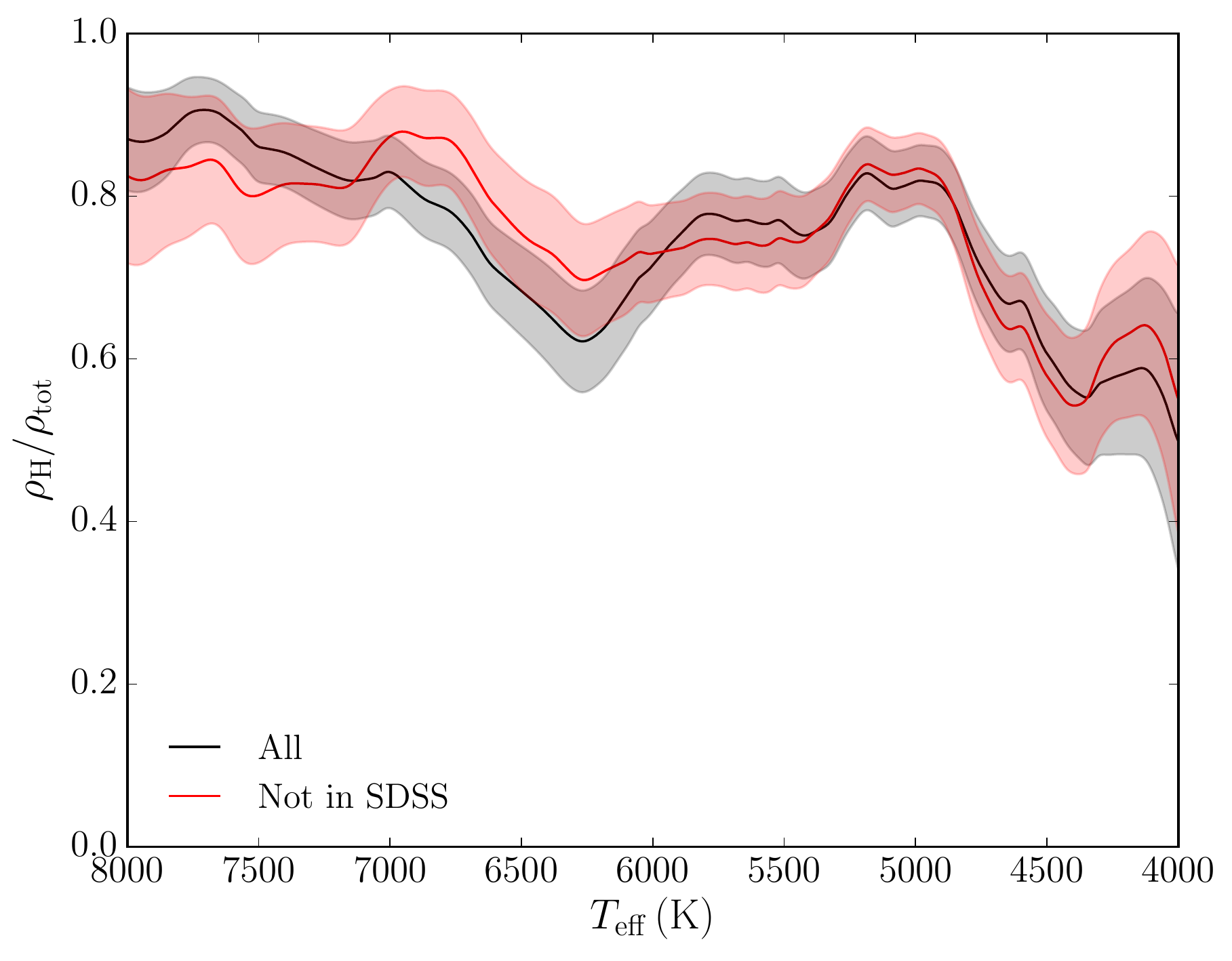}
    \caption{Fraction of hydrogen-rich white dwarfs as a function of effective
      temperature for a 500\,K moving bin. The results obtained using our whole sample
      are shown in gray and those obtained by excluding SDSS objects are show in red.}
    \label{fig:fracH_histogram_cutsdss}
  \end{center}
\end{figure}

As explained above, SDSS selection effects could lead to a bias in favor of hydrogen-rich objects
between 7000\,K and 8000\,K and a bias in favor of helium-rich objects between 6000\,K and 7000\,K
(Section~\ref{sec:evolution_biases}). To check if the decrease of $\rho_{\rm H} / \rho_{\rm tot}$
between 7500\,K and 6250\,K is solely due to this bias, we compare the evolution of the
hydrogen-rich fraction obtained using all stars in our sample to that obtained from a subsample
that excludes all SDSS objects (Figure~\ref{fig:fracH_histogram_cutsdss}).
Although both samples have different behaviors at high
temperatures, they both show a clear decrease of $\rho_{\rm H} / \rho_{\rm tot}$ between
7000\,K and 6250\,K, suggesting that this feature is not due to SDSS biases.

\begin{figure}
  \begin{center}
    \includegraphics[width=\columnwidth]{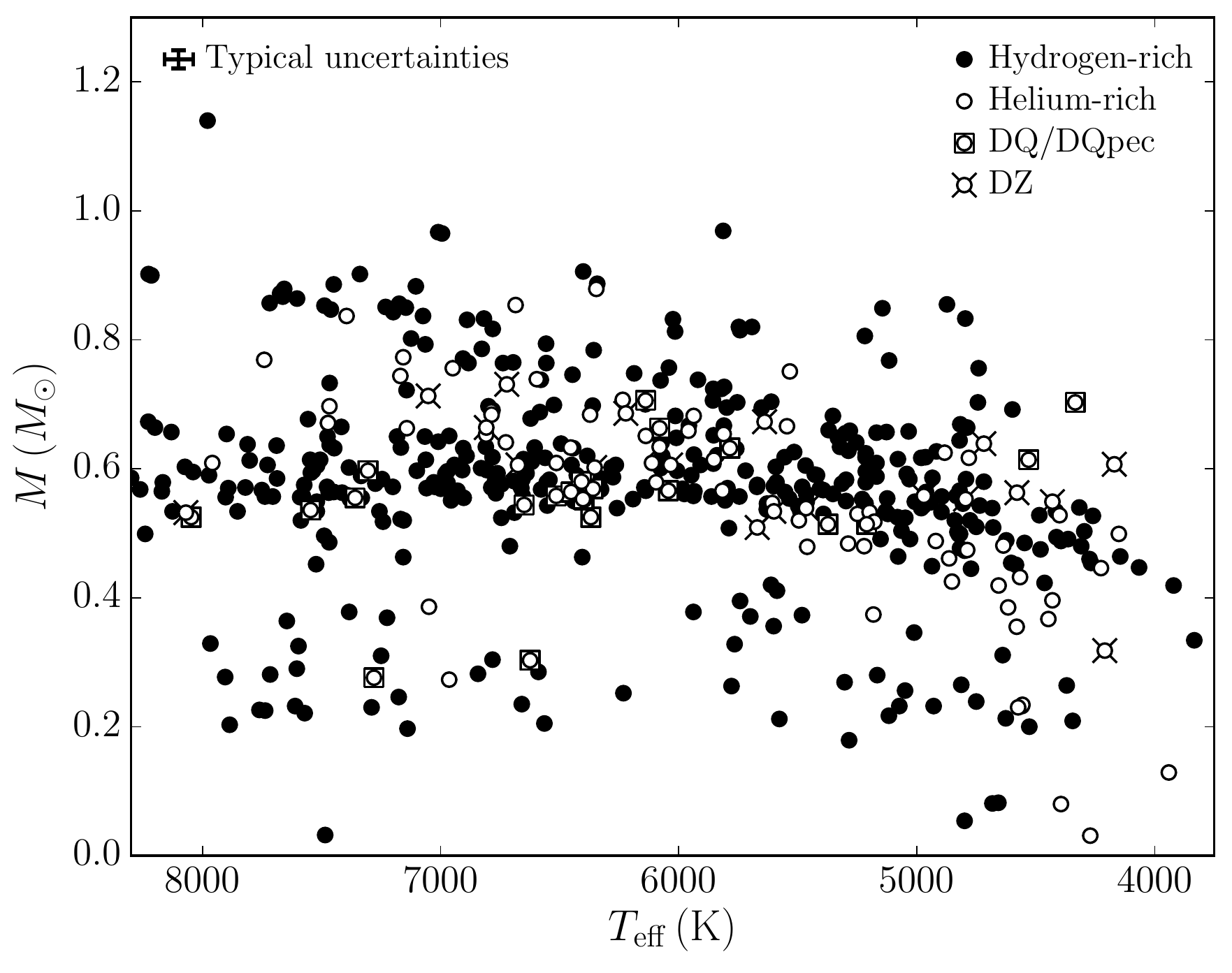}
    \caption{Mass of white dwarfs in our sample as a function of their effective temperatures.
      Typical uncertainties on $M$ and $T_{\rm eff}$ are shown in the top-left corner.}
    \label{fig:teff_mass}
  \end{center}
\end{figure}

\begin{figure}
  \begin{center}
    \includegraphics[width=\columnwidth]{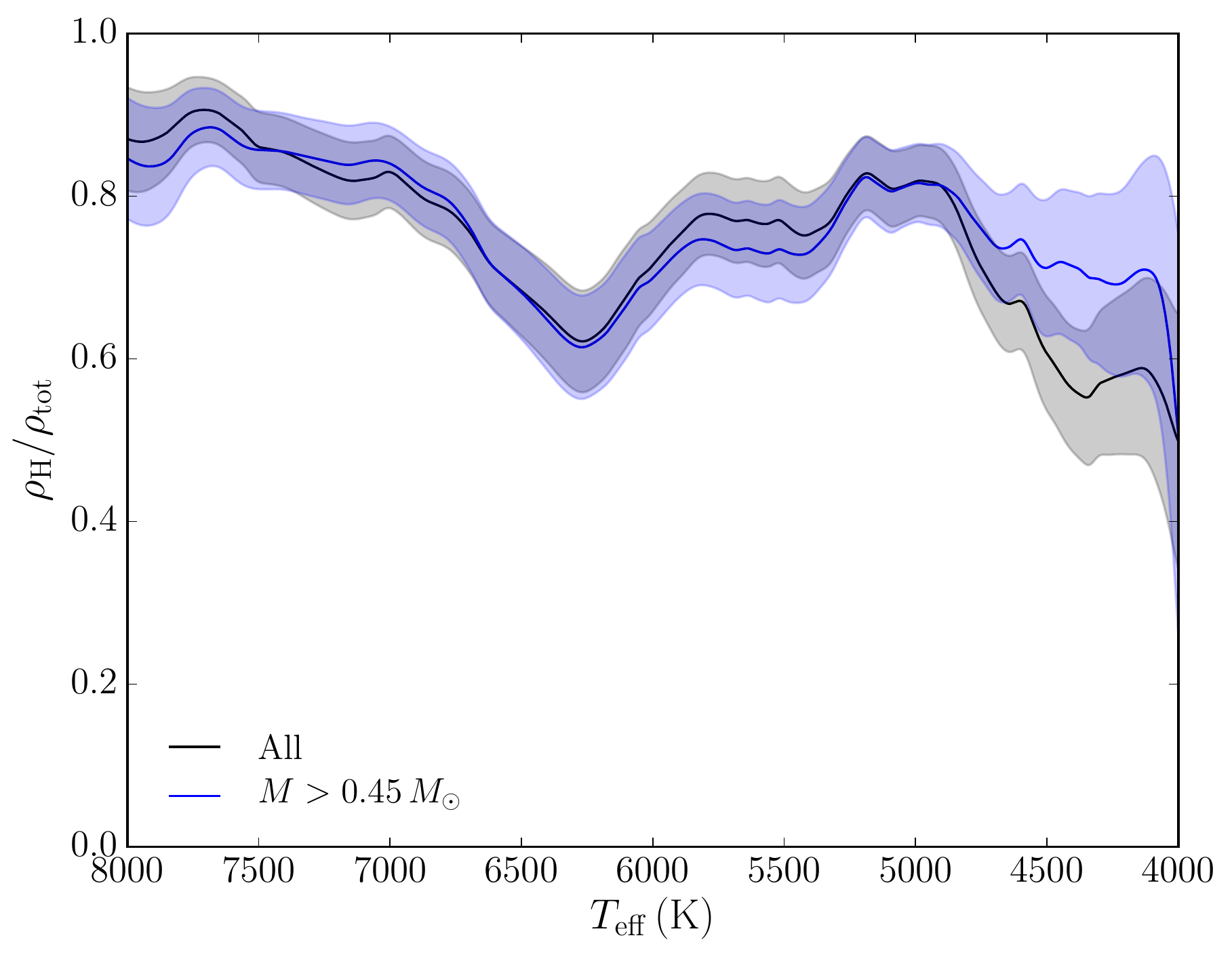}
    \caption{Fraction of hydrogen-rich white dwarfs as a function of effective
      temperature for a 500\,K moving bin. The results obtained using our whole sample
      are shown in gray and those obtained by excluding objects with $M \leq 0.45\,M_{\odot}$
      are shown in blue.}
    \label{fig:fracH_histogram_cutmass}
  \end{center}
\end{figure}

Another potential problem with our analysis of $\rho_{\rm H} / \rho_{\rm tot}$ is that it includes
several objects with a very low mass (Figure~\ref{fig:teff_mass}). Since it is impossible for a star
with a mass below $\approx 0.45\,M_{\odot}$ to become a white dwarf through single-star evolution
within the age of the Universe, those objects are likely part of binary systems
\citep{liebert2005formation,rebassa2011post}. Figure~\ref{fig:fracH_histogram_cutmass}
compares the hydrogen-rich fractions that
we find if we include and if we exclude those low-mass objects. Clearly, the evolution of
$\rho_{\rm H} / \rho_{\rm tot}$ above 5000\,K is the same for both samples, so the contamination
of our sample by binary systems cannot explain the decrease of $\rho_{\rm H} / \rho_{\rm tot}$
in the 7500$-$6250\,K range and the increase in the 6250$-$5000\,K range.

\begin{figure}
  \begin{center}
    \includegraphics[width=\columnwidth]{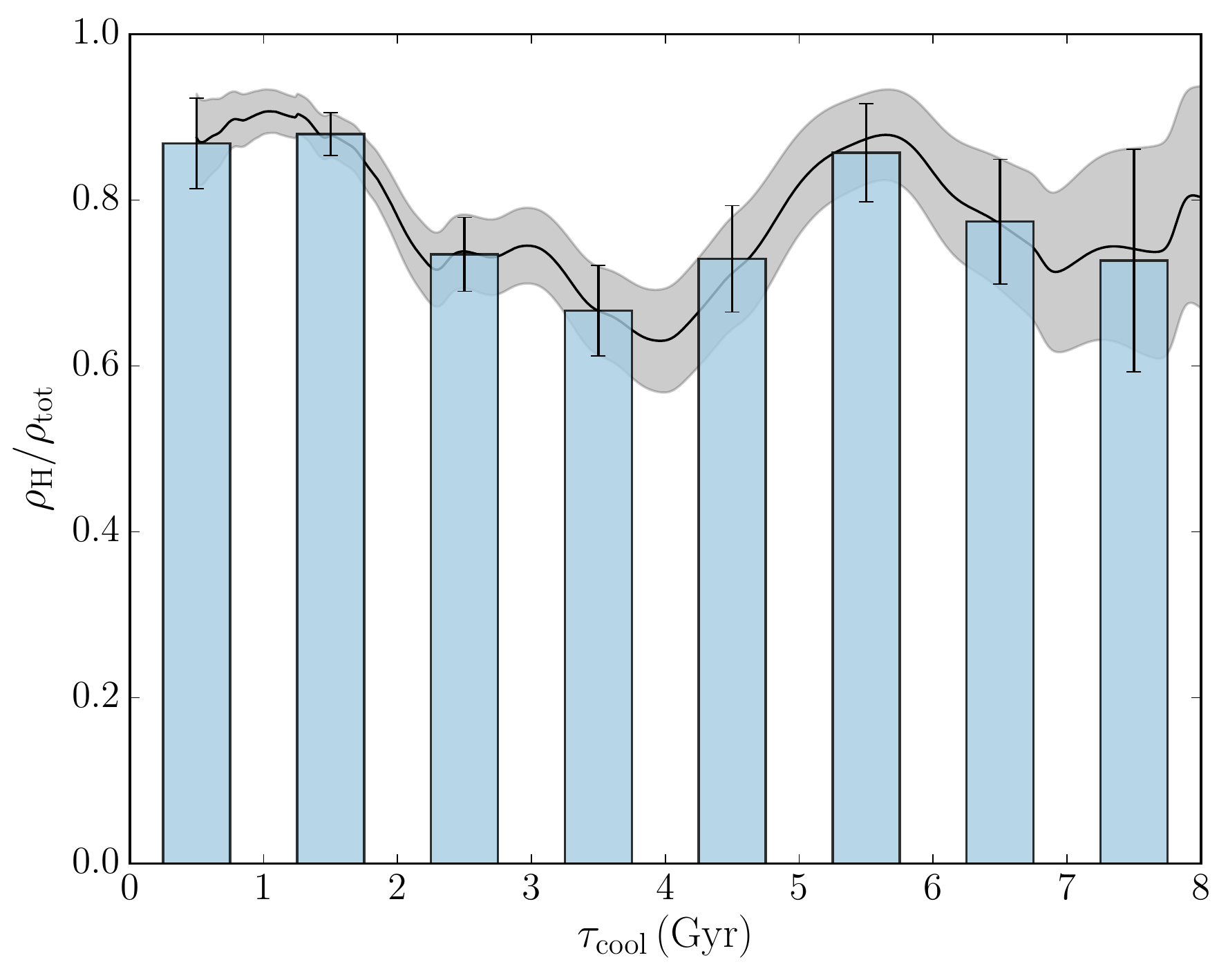}
    \caption{Fraction of hydrogen-rich white dwarfs as a function of cooling age
      for a 1\,Gyr moving bin (in gray) and for fixed 1\,Gyr bins (in blue).}
    \label{fig:fracH_histogram_age}
  \end{center}
\end{figure}

Another way to look at the evolution of the hydrogen-rich fraction is to look at
$\rho_{\rm H} / \rho_{\rm tot}$ as a function of the cooling age (Figure~\ref{fig:fracH_histogram_age}).
While the horizontal
axis is distorted compared to Figure~\ref{fig:fracH_histogram_500}, the decrease 
of $\rho_{\rm H} / \rho_{\rm tot}$ before 6250\,K and the increase down to 5000\,K are still
clearly visible.\footnote{A hydrogen-rich white dwarf with $\log g=8$ has a cooling age
  of 2.0\,Gyr when $T_{\rm eff}=6250\,$K and of 5.7\,Gyr when $T_{\rm eff}=5000\,$K \citep{fontaine2001potential}.}
This shows that the trends observed in the $\rho_{\rm H} / \rho_{\rm tot}$ vs $T_{\rm eff}$ plot
(Figure~\ref{fig:fracH_histogram_500}) are representative of the temporal evolution of
cool white dwarfs.

All things considered, the decrease of the hydrogen-rich fraction from $T_{\rm eff} \approx 7500\,{\rm K}$ to
$T_{\rm eff} \approx 6250\,{\rm K}$ and the subsequent increase down to $T_{\rm eff} \approx 5000\,{\rm K}$
appear to be real. The decrease of $\rho_{\rm H} / \rho_{\rm tot}$ is consistent with what we expect from
convective mixing. The deepening of the convection zone with decreasing effective temperature mixes the
superficial hydrogen layer with the more massive helium layer underneath, which can turn DAs into non-DAs
\citep{tassoul1990evolutionary,bergeron1997chemical,rolland2018spectral}.
However, the increase of $\rho_{\rm H} / \rho_{\rm tot}$ in the 6250$-$5000\,K range is much more intriguing.
This increase is similar to that observed at the blue edge of the non-DA gap
\citep{bergeron1997chemical,bergeron2001photometric}, albeit located at slightly lower temperatures.
No satisfactory physical explanation for this increase can be found in the literature
\citep{hansen1999cooling,malo1999physical,bergeron2001photometric} and we do not have any new
scenario to propose.

\subsubsection{Behavior below 5000\,K}
Let's now examine the decrease of $\rho_{\rm H} / \rho_{\rm tot}$ below 5000\,K. The first thing
to note is that this feature cannot be affected by SDSS selection biases, since there are very
few SDSS objects with $T_{\rm eff}<5000\,{\rm K}$ in our sample (see Figures~\ref{fig:sdss_spectra} and
\ref{fig:fracH_histogram_cutsdss}).
Secondly, we note that the decrease of the hydrogen-rich fraction becomes less obvious if we eliminate
the low-mass objects from our sample. For the blue curve of Figure~\ref{fig:fracH_histogram_cutmass},
the decline of $\rho_{\rm H} / \rho_{\rm tot}$ between 5000\,K and 4000\,K is barely significant.
Thirdly, no significant counterpart for the decrease below 5000\,K in the
$\rho_{\rm H} / \rho_{\rm tot}$ vs $T_{\rm eff}$ figure is visible if we plot $\rho_{\rm H} / \rho_{\rm tot}$
as a function of the cooling age (Figure~\ref{fig:fracH_histogram_age}). Therefore, it is not clear that 
the decrease of $\rho_{\rm H} / \rho_{\rm tot}$ below 5000\,K reflects the temporal evolution
of cool white dwarfs. Instead, the finding that helium-rich white dwarfs become more abundant below 5000\,K
may simply reflect the fact that they cool down faster than their hydrogen-rich counterparts.

In any case, our results rule out the possibility that cool white dwarfs all become hydrogen-rich as a result
of hydrogen accretion from the interstellar medium. This scenario was already challenged
by the fact that the accretion rate required for this conversion 
\citep[$\approx 6 \times 10^{-17}\,M_{\odot}\,{\rm yr}^{-1}$,][]{kowalski2006phd} is incompatible
with the limits on the accretion rates of cool helium-rich DA/DZA white dwarfs
\citep[$10^{-20}$ to $10^{-17}\,M_{\odot}\,{\rm yr}^{-1}$,][]{dufour2007spectral,rolland2018spectral}.

\section{Conclusion}
\label{sec:conclusion}
A detailed photometric and spectroscopic analysis of a homogeneous sample of 501 cool
white dwarfs was presented. Our analysis was based on a state-of-the-art model
atmosphere code that includes all the nonideal input physics required to accurately 
model the dense atmospheres of cool white dwarfs. As our models are the first to
successfully model the most challenging cool DZ white dwarfs (Paper I, II and III)
and as our analysis makes use of the largest homogeneous sample of cool white dwarfs studied to date,
our results are on firmer grounds than previous attempts at constraining the spectral
evolution of cool white dwarfs.

A satisfactory fit was found for all 501 white dwarfs studied in this work, except
for a handful of carbon-polluted objects.
We explored a few avenues to improve our fits to those stars, such as
empirically adjusting the pressure shift of the distorted Swan bands and computing
the absorption resulting from C$_2-$He collisions. 
We also revisited a peculiar class of DC stars (known as peculiar non-DAs) 
for which spectroscopic and photometric observations suggest conflicting
chemical compositions. We reported evidence that is at odds with the interpretation
according to which those objects form a real physical group.

We found that hydrogen-rich white dwarfs become less abundant when the effective temperature
decreases from 7500\,K to 6250\,K, which is likely explained by convective mixing.
From 6250\,K to 5000\,K, the trend is reversed and the fraction of white dwarfs with a hydrogen-rich atmosphere
increases. So far, no physical scenario is available to explain this observation.
Finally, at lower temperatures, we find that hydrogen-rich white dwarfs become rarer from
5000\,K to 4000\,K. This trend invalidates the scenario according to which accretion 
of hydrogen from the interstellar medium dominates the spectral evolution of cool white dwarfs.
 
\acknowledgments

We wish to thank Pierre Bergeron for many enlightening discussions on the spectral evolution of white dwarfs.
We are grateful to Mukremin Kilic for sharing with us his spectrum of SDSS J134118.68+022737.0.
We also thank Adela Kawka and Stephane Vennes for allowing us to use their VLT/FORS2 spectra of
NLTT~8733, NLTT~14553 and NLTT~57760.

This work was supported in part by the NSERC (Canada) and the Fund FRQNT (Qu\'ebec).
This work has made use of the Montreal White Dwarf Database \citep{dufour2016montreal}.

This work has made use of data from the European Space Agency (ESA) mission
{\it Gaia} (\url{https://www.cosmos.esa.int/gaia}), processed by the {\it Gaia}
Data Processing and Analysis Consortium (DPAC,
\url{https://www.cosmos.esa.int/web/gaia/dpac/consortium}). Funding for the DPAC
has been provided by national institutions, in particular the institutions
participating in the {\it Gaia} Multilateral Agreement.

The Pan-STARRS1 Surveys (PS1) and the PS1 public science archive have been made possible through contributions by the Institute for Astronomy, the University of Hawaii, the Pan-STARRS Project Office, the Max-Planck Society and its participating institutes, the Max Planck Institute for Astronomy, Heidelberg and the Max Planck Institute for Extraterrestrial Physics, Garching, The Johns Hopkins University, Durham University, the University of Edinburgh, the Queen's University Belfast, the Harvard-Smithsonian Center for Astrophysics, the Las Cumbres Observatory Global Telescope Network Incorporated, the National Central University of Taiwan, the Space Telescope Science Institute, the National Aeronautics and Space Administration under Grant No. NNX08AR22G issued through the Planetary Science Division of the NASA Science Mission Directorate, the National Science Foundation Grant No. AST-1238877, the University of Maryland, Eotvos Lorand University (ELTE), the Los Alamos National Laboratory, and the Gordon and Betty Moore Foundation.

\bibliographystyle{aasjournal}
\bibliography{references}

\end{document}